\documentclass[]{aastex631} %

\usepackage{color,soul}
\usepackage[linesnumbered,ruled,vlined]{algorithm2e} 

\def\kms{km~s$^{-1}$}
\def\HI{H{\sc i}}

\shorttitle{CHILES Deep Imaging}
\shortauthors{Dodson et al.}
\graphicspath{{./}{Figures/}}

\begin{document}

\title{CHILES VII: Deep Imaging for the CHILES project, a SKA prototype}

\author[0000-0003-0392-3604]{R. Dodson}
\affiliation{International Centre for Radio Astronomy Research, The University of Western Australia, 35 Stirling Hwy, Crawley,  Western Australia}

\author[0000-0003-3168-5922]{E. Momjian}
\affiliation{National Radio Astronomy Observatory, P.O. Box O, Socorro, NM 87801, USA}

\author[0000-0001-7996-7860]{D.J. Pisano}
\affiliation{Department of Physics and Astronomy, West Virginia University, P.O. Box 6315, Morgantown, WV 26506, USA}
\affiliation{Center for Gravitational Waves and Cosmology, West Virginia University, Chestnut Ridge Research Building, Morgantown, WV 26505}
\affiliation{Adjunct Astronomer at Green Bank Observatory, Green Bank, WV, USA}
\affiliation{Department of Astronomy, University of Cape Town, Private Bag X3, Rondebosch 7701, South Africa}

\author{N. Luber}
\affiliation{Department of Physics and Astronomy, West Virginia University, P.O. Box 6315, Morgantown, WV 26506, USA}
\affiliation{Center for Gravitational Waves and Cosmology, West Virginia University, Chestnut Ridge Research Building, Morgantown, WV 26505}

\author{J. Blue Bird}
\affiliation{Department of Astronomy, Columbia University, 550 West 120th Street, New York, NY 10027, USA}
\affiliation{National Radio Astronomy Observatory, P.O. Box O, Socorro, NM 87801, USA}

\author[0000-0002-5611-9292]{K. Rozgonyi}
\affiliation{International Centre for Radio Astronomy Research, The University of Western Australia, 35 Stirling Hwy, Crawley,  Western Australia}
\affiliation{Australian Research Council Centre of Excellence for All-sky Astrophysics in 3-Dimensions (ASTRO-3D), Australia}
\affiliation{Faculty of Physics, Ludwig-Mmaximilians-Universit\"at, Scheinerstr. 1, 81679, Munich, Germany}

\author{E.T. Smith}
\affiliation{Department of Physics and Astronomy, West Virginia University, P.O. Box 6315, Morgantown, WV 26506, USA}
\affiliation{Center for Gravitational Waves and Cosmology, West Virginia University, Chestnut Ridge Research Building, Morgantown, WV 26505}

\author[0000-0002-7679-9344]{J.H. van Gorkom}
\affiliation{Department of Astronomy, Columbia University, 550 West 120th Street, New York, NY 10027, USA}

\author[0000-0002-8799-3054]{D. Lucero}
\affiliation{Department of Physics, Virginia Tech, 850 West Campus Drive, Blacksburg, VA 24061, USA}

\author[0000-0001-9662-9089]{K. M. Hess}
\affiliation{ASTRON, the Netherlands Institute for Radio Astronomy, Postbus 2, Dwingeloo NL-7900AA, The Netherlands}
\affiliation{Kapteyn Astronomical Institute, University of Groningen, Landleven 12, 9747 AD, Groningen, The Netherlands}
\affiliation{Instituto de Astrof\'{i}sica de Andaluc\'{i}a (CSIC), Glorieta de la Astronom\'{i}a s/n, 18008 Granada, Spain}

\author[0000-0001-7095-7543]{M. Yun}
\affiliation{Department of Astronomy, University of Massachusetts, Amherst, MA 01003, USA}

\author[0000-0001-8496-4306]{J. Rhee}
\affiliation{International Centre for Radio Astronomy Research, The University of Western Australia, 35 Stirling Hwy, Crawley,  Western Australia}
\affiliation{Australian Research Council Centre of Excellence for All-sky Astrophysics in 3-Dimensions (ASTRO-3D), Australia}

\author[0000-0002-9316-763X]{J.M. van der Hulst}
\affiliation{Kapteyn Astronomical Institute, University of Groningen, Landleven 12, 9747 AD, Groningen, The Netherlands}

\author[0000-0001-5332-3784]{K. Vinsen}
\affiliation{International Centre for Radio Astronomy Research, The University of Western Australia, 35 Stirling Hwy, Crawley,  Western Australia}

\author{M. Meyer}
\affiliation{International Centre for Radio Astronomy Research, The University of Western Australia, 35 Stirling Hwy, Crawley,  Western Australia}
\affiliation{Australian Research Council Centre of Excellence for All-sky Astrophysics in 3-Dimensions (ASTRO-3D), Australia}

\author[0000-0002-9259-3776]{X. Fernandez}
\affiliation{Department of Physics and Astronomy, Rutgers, The State University of New Jersey, Piscataway, NJ 08854-8019, USA}

\author[0000-0003-1436-7658]{H. B. Gim}
\affiliation{Department of Physics, Montana State University, P. O. Box 173840, Bozeman, MT 59717, USA}
\affiliation{Department of Astronomy, University of Massachusetts, Amherst, MA 01003, USA}

\author[0000-0001-9234-1088]{A. Popping}
\affiliation{International Centre for Radio Astronomy Research, The University of Western Australia, 35 Stirling Hwy, Crawley,  Western Australia}

\author{E. Wilcots}
\affiliation{Department of Astronomy, University of Wisconsin - Madison, 475 N Charter St., Madison, WI 53706, USA}

\begin{abstract}

Radio Astronomy is undergoing a renaissance, as the next-generation of instruments provides a massive leap forward in collecting area and therefore raw sensitivity. However, to achieve this theoretical level of sensitivity in the science data products we need to address the much more pernicious systematic effects, which are the true limitation.
These become all the more significant when we consider that much of the time used by survey instruments, such as the SKA, will be dedicated to deep surveys.

CHILES is a deep \HI{} survey of the COSMOS field, with 1,000 hours of VLA time. We present our approach for creating the image cubes from the first Epoch, with discussions of the methods and quantification of the data quality from 946 to 1420\,MHz -- a redshift range of 0.5 to 0. We layout the problems we had to solve and describe how we tackled them.
These are of importance as CHILES is the first deep wideband multi-epoch \HI{} survey and it has relevance for ongoing and future surveys.

We focus on the accumulated systematic errors in the imaging, as the goal is to deliver a high-fidelity image that is only limited by the random thermal errors. 
To understand and correct these systematic effects we ideally manage them in the domain in which they arise, and that is predominately the visibility domain.
CHILES is a perfect test bed for many of the issues we can expect for deep imaging with the SKA or ngVLA and
we discuss the lessons we have learned.

\end{abstract}




\section{Introduction}
The greatest challenge in the current era, when Radio Astronomy is under-going a generational refresh of available infrastructure, is to deliver on the promise of the new instruments such as the Square Kilometre Array (SKA)\footnote{http://www.skatelescope.org} and the next-generation VLA (ngVLA)\footnote{http://ngvla.nrao.edu}.
These deliver several orders of magnitude improvements in effective collecting area -- and thus thermal sensitivity. But in many cases the images we produce currently are limited not by the theoretical sensitivity but by the systematic errors; our processing needs to be equally improved along with the infrastructure \citep[see discussions in][for example]{rioja_20}.
This is further complicated by the sheer volume of data from the next-generation instruments that we will need to process, which is increased by an order of magnitude more than the collecting area.
A large fraction of the SKA science projects are requesting many days worth of data, spanning years in some cases. In these cases it is not feasible to store the data for a final processing. 
With smaller data volumes it is possible to iteratively cycle through the raw visibilities and hunt down and remove the sources of systematic error. 
This could be done with many years worth of data combined, to improve the sensitivity to subtle effects.
In the future we will need to process our data immediately, as it can not be stored, whilst removing systematic effects to a higher accuracy than can be detected in the batch of data currently being processed. 
This will be a challenge, which we are partially exploring using the COSMOS \HI{} Large Extragalactic Survey (CHILES) project. This provides an ideal example and test case for trying out new techniques and approaches. 

In this investigation we are testing pipeline solutions for the calibrated and flagged datasets from the Karl G. Jansky Very Large Array (VLA) for the deep \HI{} survey CHILES \citep{chiles_pilot,chiles_detect}, which completed observations in 2019.
This survey aims to study the neutral atomic hydrogen (\HI) content of galaxies over 4 billion years of cosmic time, approximately 1/3 of the history of the Universe and twice the lookback time of any previous emission-line survey \citep{hipass,alfalfa,budhies}.
\HI{} is a crucial ingredient to study for understanding galaxy evolution, as it is the dominant baryonic fuel out of which stars and galaxies are ultimately made, as well as being an important tracer of galaxy kinematics.  Such surveys have previously been too expensive to carry out due to limitations in both back-end processing resources and observing time. CHILES was selected as a large program with an observing time  of 1,000 hours, to demonstrate the unprecedented capability introduced by the VLA upgrades.
Therefore the CHILES survey is one of the prime pathfinders for data processing on SKA scales,
as the data volumes and processing requirements of this project stretch the bounds of current computing capability.
The CHILES data volume matches the expected scale for the daily SKA imaging. Therefore we have used it extensively to test and understand SKA Data Processing issues. 
{The \typeout{God-like} WIDAR correlator was capable of producing commensal spectral and continuum data products. The latter leveraged the large observing time for the \HI{} line for high-sensitivity continuum polarization imaging and high-cadence transient surveys, CHILES ConPol \citep{chiles_conpol} and CHILES VERDES\footnote{https://web.pa.msu.edu/people/chomiuk/ChilesVerdes/index.html} respectively.} 
Here we present the data analysis of the CHILES spectral line dataset, with the method and the data quality metrics used.

In this study we have been exploring the issues around deep imaging, as we believe the current SKA and pathfinder strategies are sub-optimal.
The previous interferometric deep \HI{} survey, BUD\HI ES, had significantly lower spatial resolution and a narrower frequency coverage. These limited the `big-data' aspect of that project, but the imaging considerations had significant overlap with CHILES.
The big-data limitation arises from the required storage space for SKA-era spectral line observations, of which only about a days-worth of raw visibility data can feasibly be stored for processing.
Thus the current {SKA (and pathfinders)} strategy for deep spectral line imaging is to image each individual day and stack these images to produce the final deep image data product. This is theoretically acceptable for thermal-noise limited images, 
however, in the case of systematic-noise limited images, the proposed averaging of daily images will `bake-in' certain errors and the residuals will not average down. 
We have completed a full analysis of these issues and demonstrate the effectiveness of a possible solution in \citet{kristof_phd}; here we present the more traditional approach, which is still possible for this extreme example of current datasets. Our analysis of the deep CHILES imaging has demonstrated a multitude of systematic limits, which would not be possible to resolve in the image domain. We can resolve them with the full visibility dataset, as we describe here, but similar problems would be insurmountable with the current SKA approaches to deep spectral line imaging.
{Previous comparable `thousand hour' projects, such as BUD\HI ES \citep{budhies} with WSRT or contemporary projects, such as LADUMA \citep{blyth_16}
with MeerKAT, had significantly smaller datasets and also more well-behaved instruments due to either HA-DEC or Offset Gregorian mounts, respectively.
Therefore these either would not have met these issues, or will not yet have met these issues (particularly as, at the time of writing, LADUMA has only begun to combine their observations). 
For BUD\HI ES the data analysis was done in AIPS \citep{aips}, with data volumes approximately 400 times smaller. 
Systematic errors for LADUMA in averaging images from multiple epochs were identified as the key risk in their data analysis 
\citep{MeerKAT_LPR} 
and progress against these risks is not yet available.
They use a similar setup as for CHILES, i.e. the Common Astronomy Software Applications (CASA) package \citep{mcmullin2007casa} inside independent instances. 
However it would not be at all surprising that when they start to integrate up towards a thousand hours that systematic errors will appear, which will require  more sophisticated as-yet undeveloped analysis pipelines.
We also wish to mention the other great endeavor that is addressing similar systematic error challenges: the measurement of the Epoch of Re-ionization (EOR). The LOFAR and MWA telescopes, among others, are slowly resolving the issues and producing deeper and deeper upper-limits to the 21-cm signal power-spectrum, as for example discussed in \citet{patil_16}. 
As yet the accumulated data is not yet sufficient to provide a significant limit, but \citet{li_19} achieved a 2$\sigma$ upper limit of $\Delta^2_{21}<49^2$mK$^2$ at z$\approx$6.5 and k$\approx$0.59\,h\,c\,Mpc$^{-1}$ using 40\,hours of Phase II MWA data, and \citet{mertens_20} reached a best 2$\sigma$ upper limit of $\Delta^2_{21}<73^2$mK$^2$ at z$\approx$9.1 and k$\approx$0.075\,h\,c\,Mpc$^{-1}$ with 141\,hours of LOFAR data.
When these projects are able to integrate a thousand hours of data they expect to detect the EOR signal.

Finally we should note that there are a good number of wide-field continuum deep surveys that reach much deeper sensitivities; towards $\sim\mu$Jy/beam. 
As the partner continuum survey, CHILES ConPol \citep{chiles_conpol} has not yet published the full analysis of their particular imaging issues, we focus on the  `VLA Frontier Fields Survey' 
\citep[][]{heywood_21}. 
The continuum surveys do not have the data volume issues, as the channel-widths are much coarser, and all the data can be retained for multi-pass processing.
The VLA Frontier Fields Survey did not report having to deal with out-of-beam sources, and managed directional dependent spectral instrumental effects by independently deconvolving the full 2GHz bandwidth in four sub-bands.
We believe that the same issues we meet will be in their data, but suppressed by averaging over the variations. In which case improvements should be possible. 
}

\section{The CHILES observations}

The CHILES survey is run on the VLA, which is a 27 antenna array. 
The new, upgraded front-end (wide-band L-band receivers) and back-end (the WIDAR correlator) \citep{evlamemo_152,evlamemo_204}, provided through the Expanded Very Large Array project \citep{vla_backend}, can now provide instantaneous coverage for \HI{} spectral line observing between $\sim$950 and $\sim$1430-MHz on the sky (for the CHILES case we used 15 spectral windows (SPW) of 32\,MHz, giving a total of 480\,MHz in each session). 
The different daily observations are dithered in frequency by $\pm$5\,MHz to smooth out the edges of the spectral windows and ensure that there are no gaps in frequency coverage.
The antennas (being 25m in diameter) have a field of view (FoV) of about 0.5 degrees ($\sim$2,000 arcseconds) on the sky at 1.4\,GHz, and were pointed at RA(J2000) 10h01m24s Dec(J2000) 2d21m00s. 
The array configuration is VLA-B, which has 11\,km baselines and a typical beam size of $\sim$5\arcsec{}$\times$6\arcsec{} at 1.4\,GHz and at this declination, with the Briggs robust weighting of 0.8 (as implemented in CASA). 
The recorded data is in 15.625 kHz-wide channels (representing about 3\,\kms{} at the rest frequency of the \HI{} line being observed). 
Therefore there are 351 baselines and a little less than 31,000 channels per polarization product to be consumed. The time integration is limited by the maximum baseline and the field of view (i.e. the antenna size), and is set to 8 seconds. Typical daily split-out measurement sets are about 100\,GB on the target, per hour of observation. Furthermore the required image parameters are also large; the final full field of view is 1201x1201 pixels at critical sampling in the image plane per spectral channel, but we need to form the image beyond this final limit to clean out more distant sources. 

These datasets are therefore much larger than those normally analyzed, and therein lies the challenge. 
We will eventually fully process five epochs of observing separated by approximately 15 months between each two consecutive epochs. We report here on the data in the initial full-range cube, which is from only the first Epoch. Sixteen channels have been averaged together for 250kHz channel width for these initial images, which is equivalent to 50\,\kms{} at {\it z}=0.
Eventually we will produce the 1,000 hour image cubes with a range of
frequency resolutions and spatial pixel sizes (down to $\sim$10\,\kms{} and 1\,\arcsec) and include post-detection postage stamp cubes.
This internal cube will be used to verify the method reported here, by the CHILES science teams.
The first Epoch was of 178 hours in total, broken into 42 days of observing, spanning 
2013 November 9 through 2014 January 21, with each day's observation being between 1 and 6 hours long. 
This range of  observing times was to allow dynamic scheduling of the observations, and the subsequent epochs had much more uniform scan lengths of about 6 hours.
Table \ref{tab:params} summarises the observational parameters and the current status of processing.

\begin{table}[]
    \centering
\begin{tabular}{|l|l|}
\hline
Array configuration & VLA-B \\
Bandpass and flux density scale calibrator & 3C286 \\
Complex Gain calibrator & J0943–0819 \\
Frequency coverage (MHz) & 946 -- 1420 \\
Redshift range & 0.50 -- 0.00 \\
Synthesized beam (arcsec) & 8.6$\times$7.4 -- 6.1$\times$5.0\arcsec{} \\
Imaged Frequency channel width (kHz) & 250 \\
Velocity resolution (km s$^{-1}$) & 79 -- 52 \\
Imaged Pixel Size & 3\arcsec{}, 2\arcsec{} \\
Spatial resolution (kpc) & 0.35 -- 50.0\\
\hline
\end{tabular}
\begin{tabular}{|c|c|c|c|c|c|}
\hline
Dataset & Epoch 1 & Epoch 2 & Epoch 3 & Epoch 4 & Epoch 5 \\
\hline
Observation date & 2013--2014 & 2015 & 2016 & 2017--2018 & 2019 \\
Integration (h) & 178 & 207 & 178 & 228 & 177 \\
No. Days (ndays) & 42 & 36 & 49 & 48 & 28 \\
rms noise & 150--50 & 150--40 & - & - & 150--50\\
($\mu$Jy beam$^{-1}$ channel$^{-1}$) & & & & &\\ 
Typical 1$\sigma$ N\HI{} & 6.4--1.4 & 6.4--1.1  & - & - & 6.4--1.4\\
($\times$ 10$^{20}$ channel$^{-2}$ch$^{-1}$) & & & & & \\
\hline
    \end{tabular}
    \caption{{Table of relevant imaging parameters, for both the entire survey and for the individual epochs, where available.}}
    \label{tab:params}
\end{table}

\section{CHILES pipelines}

We briefly mention the calibration and flagging pipelines, before focusing on the topic of this paper, the imaging pipeline and conclusions from it.
Epoch 1 was calibrated using a modified version of the NRAO VLA continuum pipeline and processed on the computers at NRAO in Socorro \citep[see][]{chiles_far}.  Since this was  in the early days of the correlator there were considerable advantages in having frequent contact with NRAO staff during the processing.   
Since the first (2014) and last (2019) observations our flagging strategies have continued to evolve significantly (Pisano et al, in prep).
For the data presented here the flagging was based on the NRAO pipeline, which was painfully slow as it required a great deal of human interaction. 
The improvement has been to introduce masking based on experience, which has reduced the requirement for human input. This has allowed us to accelerate the progress and to expect to complete the flagging shortly. 
We have compared the Epoch 1 data from the old and the new pipelines to show that the results are essentially identical.
Thus for the purpose of illustrating the imaging strategy for CHILES in this paper the old flagging of Epoch 1 is entirely satisfactory.

Radio Frequency Interference (RFI)  is  a serious problem especially in the 1170 to 1300\,MHz range\footnote{https://science.nrao.edu/facilities/vla/observing/RFI/L-Band}. RFI tends to be temporally limited (impulsive broad spectrum emission from, for example, powerlines), or spectrally limited (narrow band emission from clock harmonics or communication channels).
The various GPS constellations broadcast in this frequency range and these, when the satellites are in the sidelobes of the primary beam, appear to be the most problematic for \HI{} surveys.
However, \citet{chiles_far} demonstrated that, even in the highly contaminated region between 1200 and 1300\,MHz, there are frequency intervals where galaxies can be detected. In our Epoch 1 data about 15\% of the channels are not usable due to RFI. 

\begin{algorithm}[ht!]
\BlankLine
\KwIn{Calibrated and Flagged target datasets}
\KwOut{The image cube}
\BlankLine
\SetAlgoLined
\Begin{
\SetKwProg{Pn}{Function}{:}{\KwRet{MS$^{\prime}_{{\rm day}, \Delta\nu}$}}
\Pn{\textup{Split Step} Separate MS into working sub-bands MS$_{\Delta\nu}$}{
  \ForEach{day $\in$ days}{
  \ForEach{${\Delta\nu}$ $\in$ $\nu$}{
        MS$_{{\rm day}, \Delta\nu}$ $\gets$ {\sc mstransform}(MS$_{day}$) \tcp*{select the correct Doppler-shifted channels.}
        MS$^\prime_{{\rm day}, \Delta\nu}$ $\gets$ {\sc uvsub}(MS$_{{\rm day}, \Delta\nu}$) \tcp*{Subtracts the in-beam SI model per SPW}
        \ForEach{outlier $\in$ outliers}{
                MS$^{\prime \prime}_{{\rm day}, \Delta\nu}$ $\gets$ {\sc fixvis}(MS$^\prime_{{\rm day}, \Delta\nu}$) \tcp*{Rotate to the center of outlier model}
                \ForEach{time $\in$ HA range}{
                  {\sc im.select}(MS$^{\prime \prime}_{{\rm day}, \Delta\nu}$, time) \tcp*{Select the data in time range}
                  {\sc uvsub}(MS$^{\prime \prime}_{{\rm day}, \Delta\nu}$, time) \tcp*{subtracts the out-of-beam cube model per SPW}}
                MS$^{\prime}_{{\rm day}, \Delta\nu}$ $\gets$ {\sc fixvis}(MS$^{\prime \prime}_{{\rm day}, \Delta\nu}$) \tcp*{Rotate back to the phase center of observations}
                  }}}
    }
    \BlankLine
\SetKwProg{Pn}{Function}{:}{\KwRet{$I_{\Delta\nu}$}}
\Pn{\textup{Image Invert Step} Form ${\Delta\nu}$ image cube from each separate sub-band ${\Delta\nu}$}{
      \ForEach{${\Delta\nu}$ $\in$ $\nu$}{
        $I_{\Delta\nu}$ $\gets$ {\sc tclean}($\Sigma^{\rm days}$ MS$^{\prime}_{{\rm day}, \Delta\nu}$) \tcp*{Image the residual uv data}
                  }}
                  
    \BlankLine
\SetKwProg{Pn}{Function}{:}{\KwRet{$I_{\rm chunk, line}$}}
\Pn{\textup{Image Concatenate Step} Combine $I_{\Delta\nu}$ image cubes in groups of 6 and steps of 5}{
      \ForEach{chunk $\in$ $\Delta\nu$}{
         $I_{\rm chunk}$ $\gets$ {\sc ia.imconcat}($\Sigma^6_0 I_{\Delta\nu}$) \tcp*{Combine subbands into chunks}
         $I_{\rm chunk, line}$ $\gets$ {\sc imconsub}($I_{\rm chunk}$) \tcp*{subtract continuum}
                  }}
                  }
\caption{The logical flow of the CHILES imaging pipeline, with all the crucial processing steps, as implemented with the github project {\it aws\_chiles02}\label{alg:pipeline}}
\end{algorithm}

\begin{figure}
	\plotone{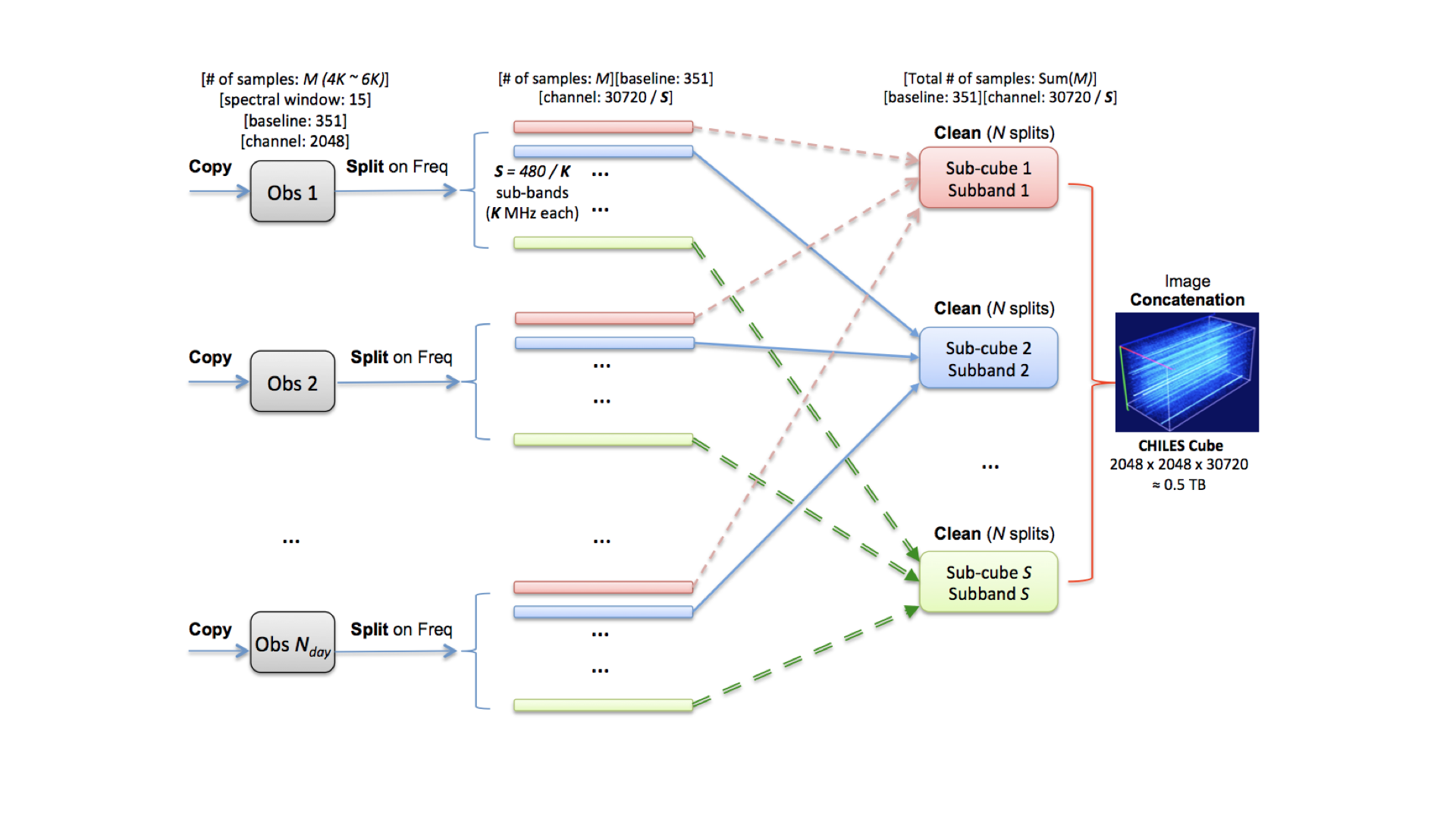}
    \caption{A schematic of the CHILES workflow, from \citet{chiles_env}. The individual days of observations are split into many sub-bands, which are then combined across the days to allow fully parallel image processing before combination into the final data product.
    \label{fig:workflow}}
\end{figure}

\subsection{Data Imaging}
\label{sec:drt}
The imaging pipeline is controlled by the Data Activated Liu Graph Engine (DALiuGE\footnote{https://github.com/ICRAR/daliuge}) \citep{daliuge_wu}, which is a workflow graph-execution framework specifically designed to support very large scale processing graphs (i.e. the connection and distribution of tasks in a parallel pipeline). It is being developed for the massive work-flow management challenge of the SKA.
This was discussed in our paper on three very different computing environments \citep{chiles_env}, where we laid out the basic steps and measured the compute requirements for these tasks, and is shown in Figure \ref{fig:workflow}. The more detailed workflow is presented in Algorithm \ref{alg:pipeline}.
The data processing is essentially the same as before, {but we now have improved our analysis and identification of the key steps for achieving the thermal limits, within the previously discussed processing pipeline. These are the focus of this paper, as we believe we are the first deep wideband multi-epoch spectral line survey to confront and solve these challenges.}
We have used Amazon Web Service (AWS) for all of the imaging pipeline, apart from the continuum modeling and some of the statistical analysis.
The CHILES imaging pipeline is accessible on github\footnote{github.com/ICRAR/aws-chiles02}. 
The  data-reduction tools we used for imaging are exclusively from CASA 
Version 5.7. 
The data provided had been calibrated and the target field selected, therefore the remaining operations to be performed in the \HI-data reduction are: to copy the data to and from the most suitable S3 cloud storage, to split the data into manageable sizes whilst allowing for the station Doppler shifts of that day, selecting specific frequency ranges and to Fourier-invert the observed data (taken in the reciprocal of the image domain) into a 3-D image cube (these dimensions being Right Ascension, Declination and Velocity). Figure 1, from \citet[][]{chiles_env}, illustrates these steps. 
Additionally one should deconvolve the image to correct the initial `dirty' image for the spatially extended point spread function (PSF), which is the Fourier Transform of the sampling function in the u,v visibility plane where interferometer measurements are made.
After the deconvolution the PSF is replaced with a compact Gaussian 
with Full Width Half Maximum (FWHM) chosen to match the resolution of the original PSF, thus forming a ‘clean’ image.
See \citet{TMSv3} for a full discussion of these concepts. 
Traditionally one would read all the data files simultaneously and invert to produce an image cube, but this is not possible as the task {\sc clean} \citep{Hogbom:1974uf} fails due to the extreme size of the input datasets.
An alternative approach, and currently the one proposed for the next-generation deep surveys, is to image each individual day of data and then combine the images. This has been shown a) to be significantly less accurate than imaging the full data \citep[see][in prep.]{kristof_deep} and b) will be very sensitive to systematic errors, which are endemic as we demonstrate in this paper.
These arise, roughly, both from the non-linear nature of the CLEAN deconvolution 
and from the summation of images with a single weight per image. The images themselves have different local weights in the uv-domain and potentially non-Gaussian error distributions.
Therefore it was essential to find an approach that would allow us to image the full visibility dataset. 
%
Our approach was to split the input data into smaller frequency ranges, or sub-bands, of 4-MHz and perform {\sc tclean()} on all of the days simultaneously, but with fewer channels. As each channel image is nominally independent this then would allow us to take advantage of the natural parallelism of the problem, and convert the challenge to a data management problem controlled by DALiuGE.
{The sub-band width of 4-MHz was selected as a compromise between parallel performance, dominated by the efficiency as a function of the number of channels formed in {\sc tclean()}, and the effects of the limited frequency span on the spectral noise, as a consequence of the robustness parameter. As such it will be sensitive to the compute hardware and performance, and the size of the data challenge.}
Other imaging software options exist (e.g. WSClean or ASKAPSoft), but (as ever) the software that works best for a particular instrument tends to be that developed by the managing institution. These alternatives would still have the same challenges in the volume of data to process. 
We pre-select the data from the input data by splitting it with {\sc mstransform}().
We no longer apply the correction to the required Barycentric frame such that the data can be added coherently in velocity when forming the images, as was done in \citet{chiles_env}, but merely ensure the frequency range selected accounts for the Doppler shift of the observatory. It was found that it was best to apply the frame correction in the {\sc tclean()} task.

Since much of the \HI{} is faint and extended, the \HI{} sources discovered will need cleaning with carefully selected masks. This is an iterative procedure, where the sources first need to be found after continuum subtraction and  masks are created using different smoothing kernels. Thus in the formation of these cubes we have {\it avoided} the deconvolution of any \HI{} emission, leaving this for the later analysis.
For this reason we have focused particularly on accurate subtraction of continuum models from the visibilities before imaging. 
The models were formed from continuum imaging of channel-averaged datasets on local computing infrastructure, as these were simpler to interact with. Data for models of sources far from the phase center had to be phase rotated before averaging, which has consequences for the model subtraction. This is further discussed in Sections \ref{sec:in_model} and \ref{sec:out_model}.
Finally, in the deconvolution we limit the cleaning to regions in which significant continuum flux has been detected and outside the nominal field of view.

We use the CASA task {\sc tclean()} in parallel on the frequency-split data, combining the many days into images for that frequency range. In these investigations we have only lightly deconvolved the images (\texttt{niter=2000}) over a few major cycles, after the continuum subtraction and only in those regions around where continuum emission has been subtracted (except when we investigated the residual noise as discussed in Section \ref{sec:resid_day}). This is acceptable as the majority of the continuum flux is subtracted using models made from the whole observing epoch. The field imaged is more than double that selected for final processing, being 4096 pixels of 3\arcsec{} below 1200\,MHz and  2\arcsec{} above. Future cubes will have finer pixel sizes.
 
We found that with the limited frequency ranges we were using, the noise levels per channel were very sensitive to the weighting scheme. For example uniform weighting (or Briggs weight -2) causes large increases of noise at the edges of the sub-cube. This is due to the implementation of the Briggs' scheme in casapy, which depends on the total data selected not just the data channel being imaged. In this case, where we have a limited input frequency span of data points from which to derive the weights, the edge data displays enhanced noise. This has been resolved, since CASA 5.5, with the {\sc perchanweightdensity} parameter. Future analysis will include the setting of this. 
However, with natural weighting (equivalent to Briggs weight +2) this does not occur, and for any Briggs weight greater than $\sim$1 it is only marginally detectable even without the setting of {\sc perchanweightdensity}. 
The imaging step, which dominates the processing time, takes about 12 hours per 4-MHz sub-band on a moderately powerful instance; on AWS we use the 16 core instance i3.4xlarge, which is typically 40c per hour on the AWS spot market \citep[See][for details of how we use the spot market]{daliuge_wu}.

The final operation is to combine the center 1201 pixels of the individual small spectral cubes into the required full sized spectral cube and subtract any residual continuum emission. 
The task {\sc ia.imageconcat()} was used to combine the individual 4-MHz sub-band image cubes into the larger 24MHz image cube chunks.
For continuum subtraction we removed a linear fit to the flux in each image pixel over 24\,MHz. 
This is wide enough to avoid removing even the largest \HI{} line features.
These 24\,MHz chunks were stepped every 20\,MHz, to ensure that there was an overlap and to ensure a smooth piece-wise model subtraction over larger frequency ranges. 
This was performed in the task {\sc imcontsub()}, ignoring channels with RMS greater than four times the median RMS across the channels; channels that are greater than ten times the median are zeroed in the final cube. 
The final product is 25 cubes, with 1201$^2$ pixels, a total of 2,400 channels and a single Stokes, resulting in a total of $\sim$40GB when using 4 bytes per voxel. 
These are served to the collaboration via AWS or other cloud services.

\subsection{Systematic effects on the cubes}
We expected that there would be systematic effects that would have prevented reaching the thermal-noise limits from the deep images, and this was certainly confirmed.
Three categories of systematic effects that appeared in the CHILES data are discussed below.

\subsubsection{In-Beam sources}\label{sec:in_model}
\begin{figure}
    \centering
    \setlength{\unitlength}{\textwidth}
    \begin{picture}(1.0,0.37)
    \put(0.02,0.35){a}
    \put(0.5,0.35){b}
    \includegraphics[width=0.45\textwidth]{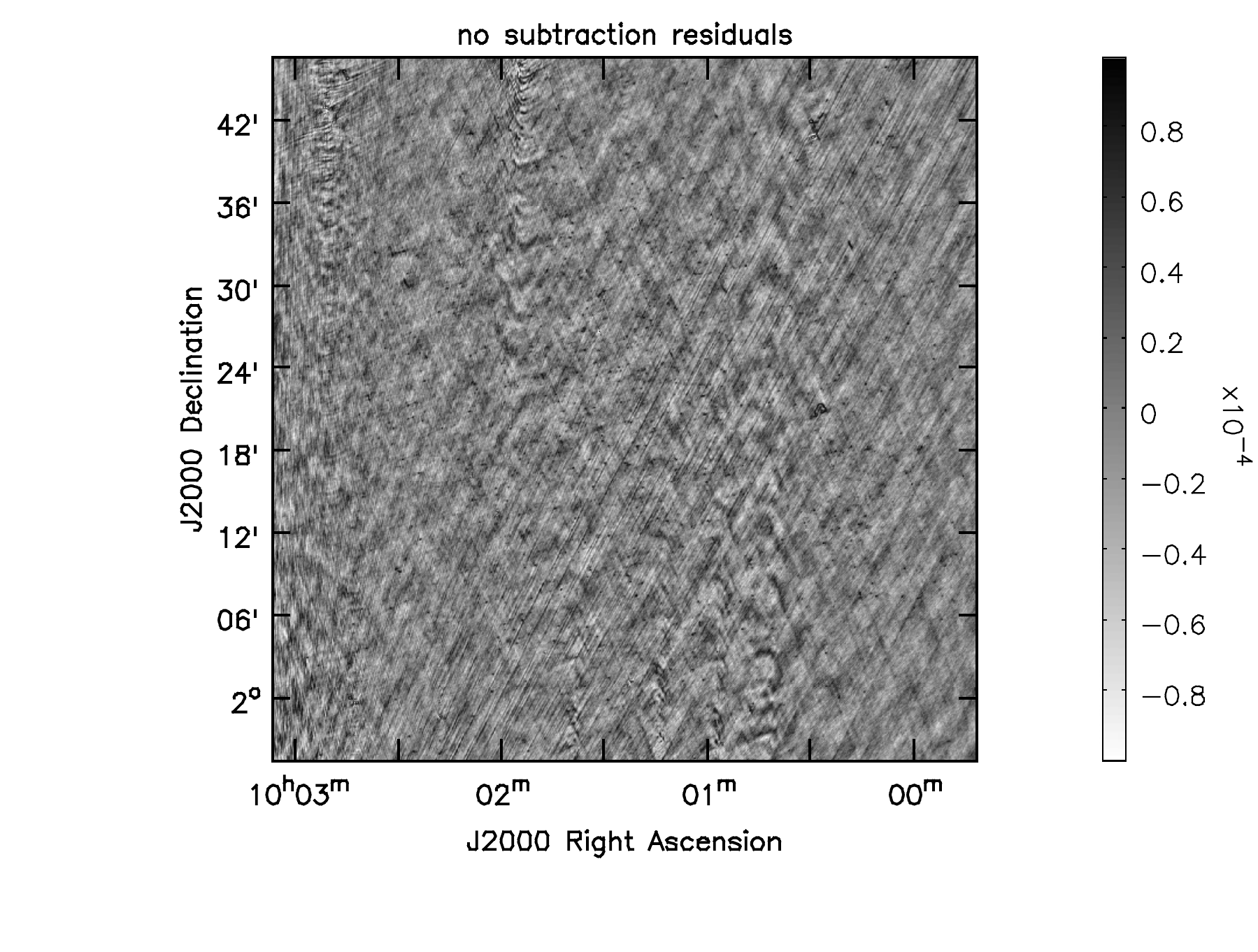}
    \includegraphics[width=0.45\textwidth]{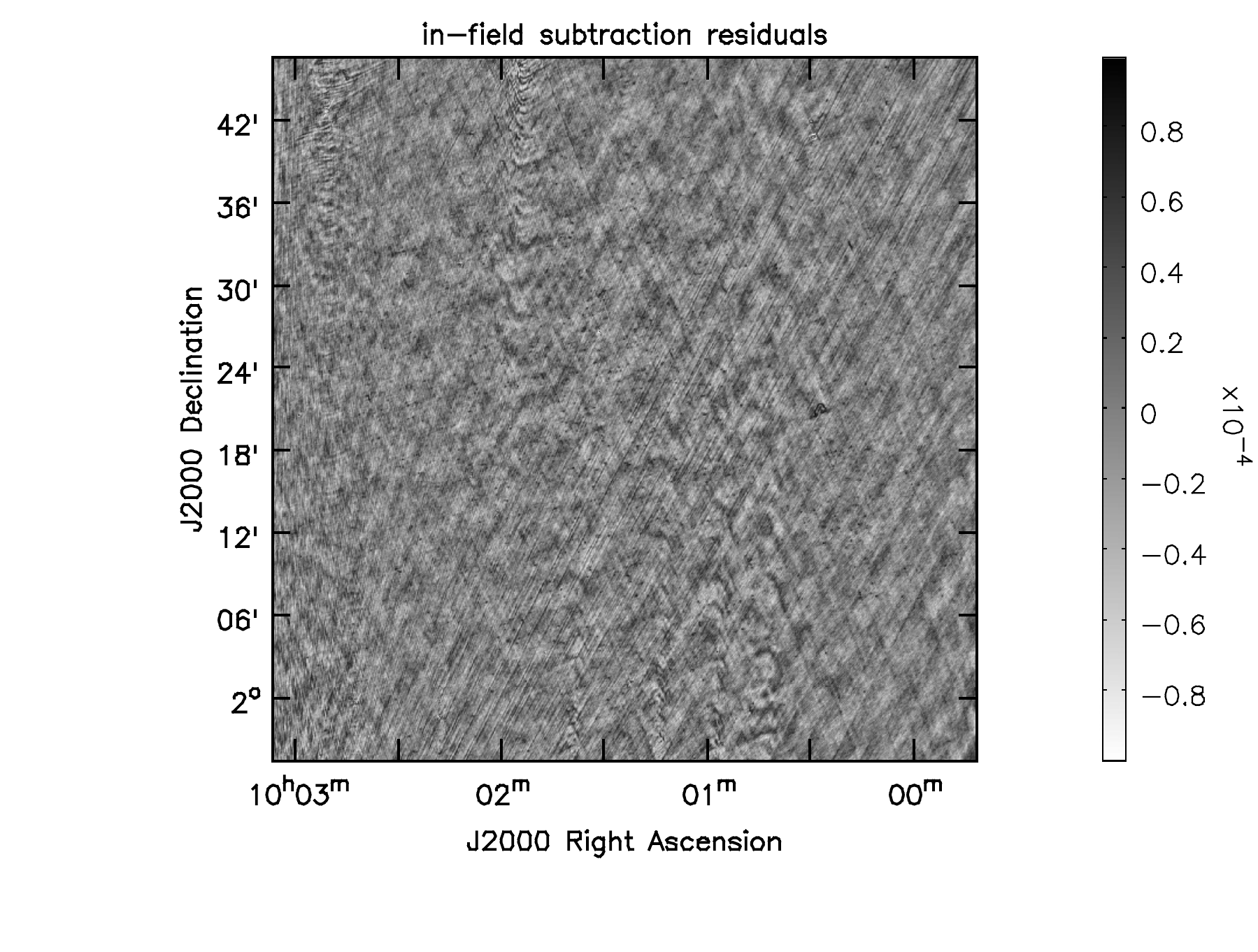}
    \end{picture}
    \begin{picture}(1.0,0.37)
    \put(0.02,0.35){c}
    \put(0.5,0.35){d}
    \includegraphics[width=0.45\textwidth]{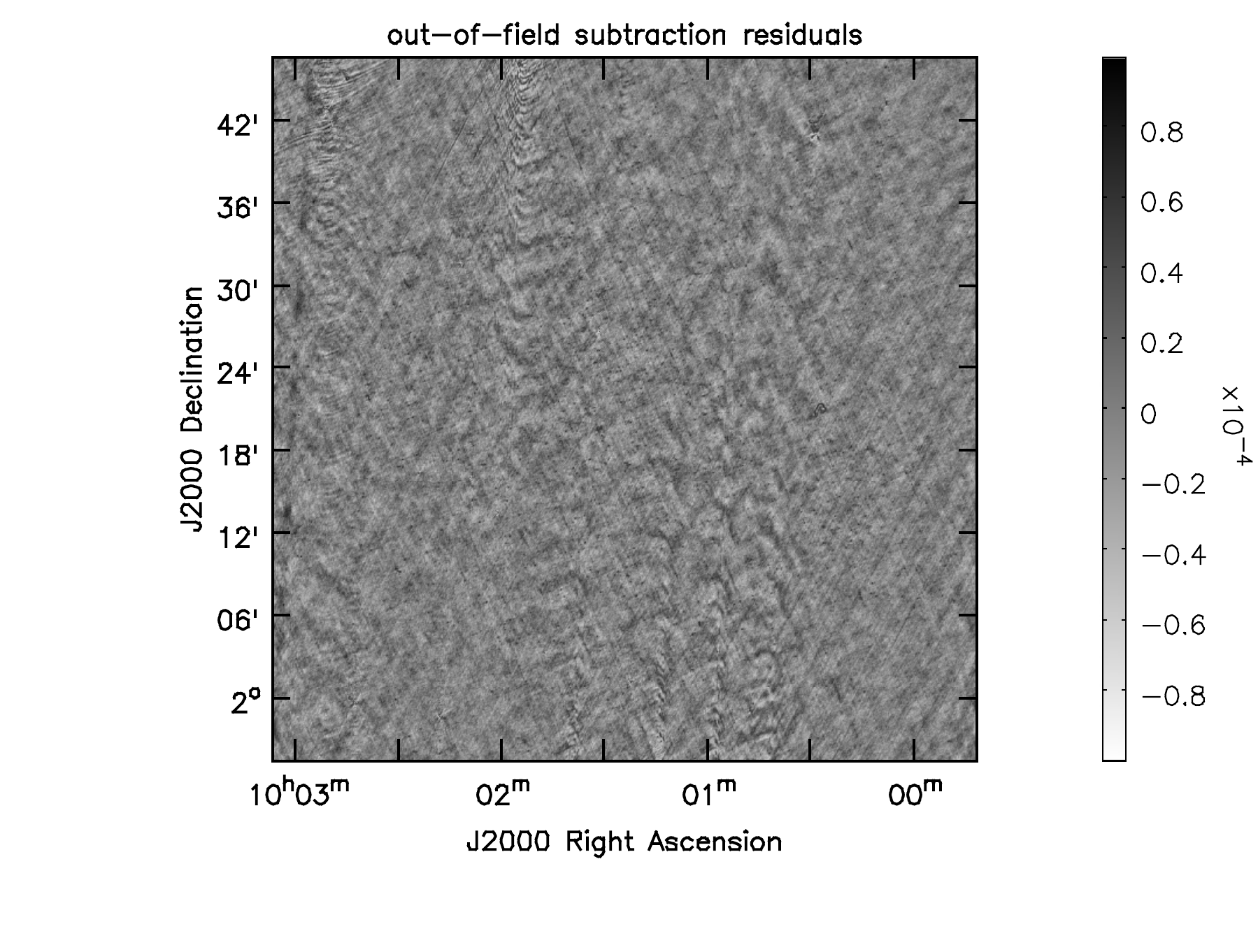}
    \includegraphics[width=0.45\textwidth]{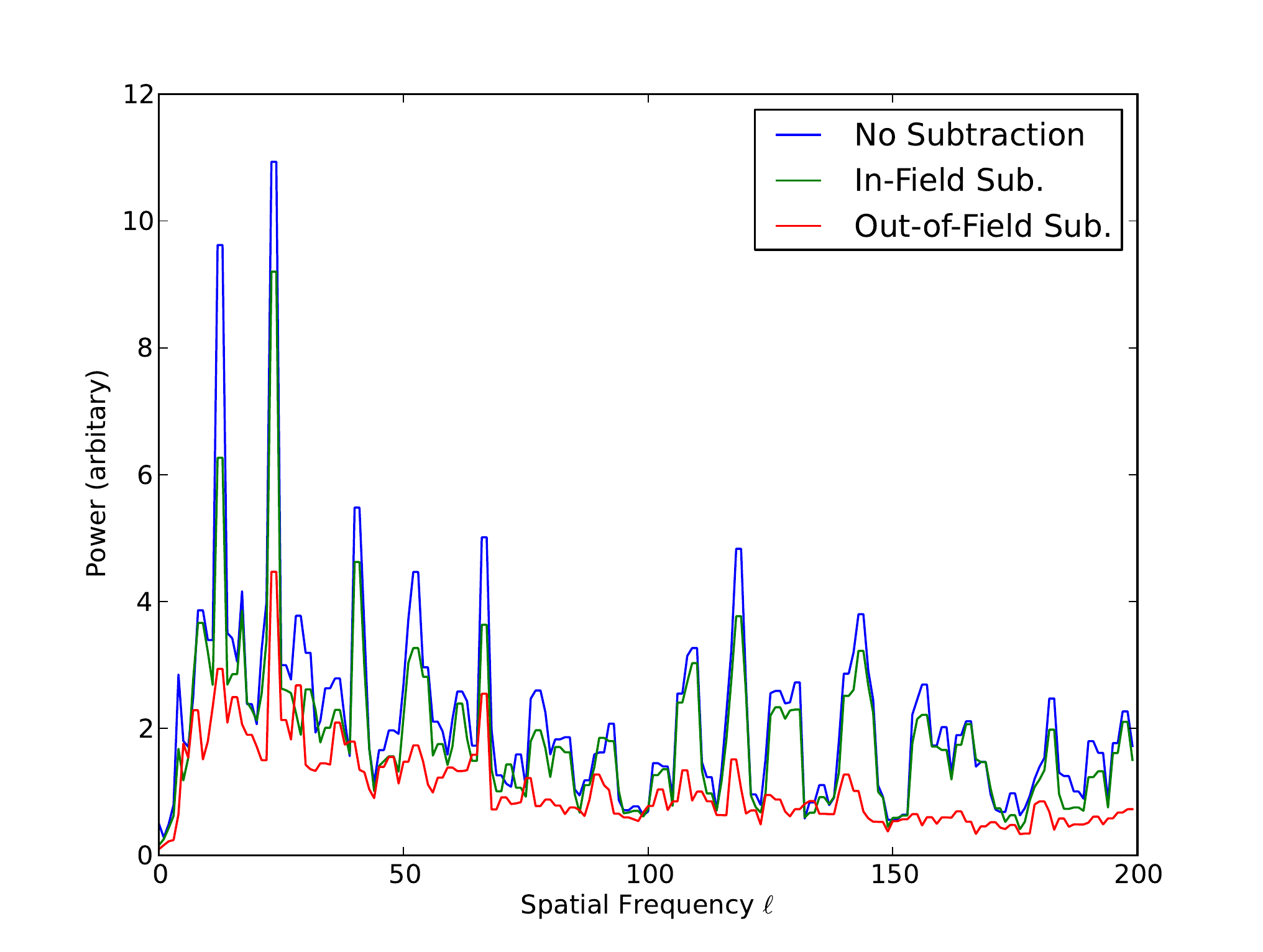}
    \end{picture}

    \caption{Continuum residual image of the CHILES field from SPW 1 (994\,MHz), scaled between $\pm$100$\mu$Jy/beam as shown with the color bar, ({\em a}) without any continuum source subtraction, ({\em b})  with subtraction of in-beam sources and ({\em c}) with in- and out-of-beam source subtraction. 
    The final plot shows the maximum peak of the radial absolute values of the FFT of the images for each stage of uv-subtraction. The suppression of the large scale structure due to the subtraction of out-of-beam sources is clear, with the multiples of the $\sim$5\arcmin{}-scale ripples significantly suppressed.
    \label{fig:in_beam}}
\end{figure}

In-beam sources were modeled with a single spectral index over the SPW. As the SPW windows were dithered, there is a $\pm$5\,MHz overlap in the spectral windows over the observations.
Therefore the models generated per SPW also have an overlap, which ensured that the piece-wise models were continuous in frequency. 
Figure \ref{fig:in_beam}{\em a} shows an image of the field of view of CHILES for SPW 1 (centered on 994\,MHz) without continuum subtraction, whilst {\em b} shows the effect of subtraction of the in-beam models, for that SPW.
These highlight the residual ripples from the outlier sources discussed in the next section. 
What constitutes an `outlier' is complicated by the large frequency range and thus the large change in the beam width over the dataset (the FWHM changes from $\sim$29$^\prime$ to 43$^\prime$ from the top to bottom of the observed frequency range). We have used the ad-hoc definition of `sources that are not well behaved' in the sense that their residuals are clearly detectable in the \HI{} cubes, which corresponds to strong sources around and beyond the FWHM.
These effects are less significant at the higher frequencies, because of the smaller FoV. 

\subsubsection{Out-of-Beam sources}\label{sec:out_model}
\begin{figure}
    \centering
    \includegraphics[width=0.3\textwidth]{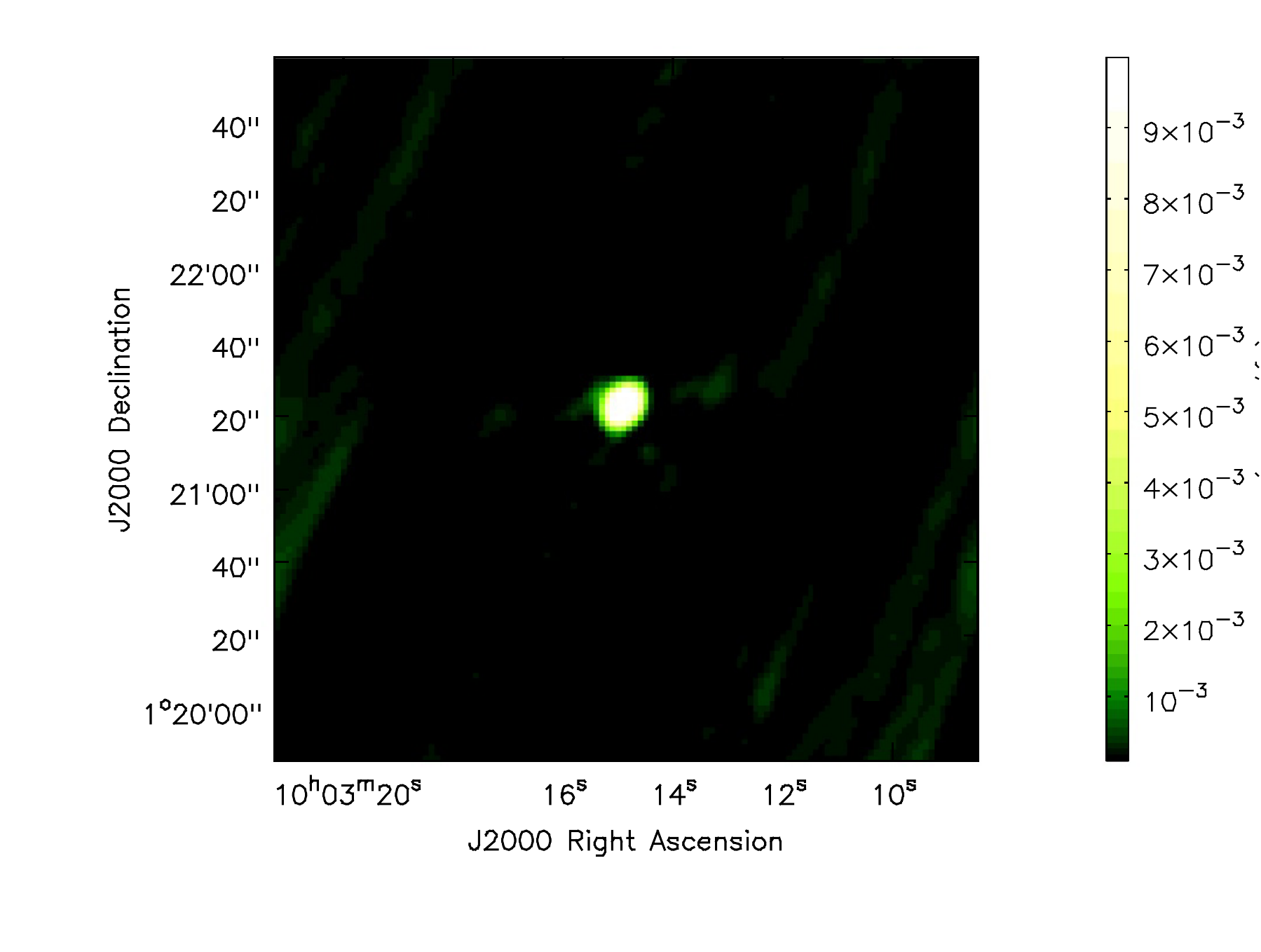}
    \includegraphics[width=0.3\textwidth]{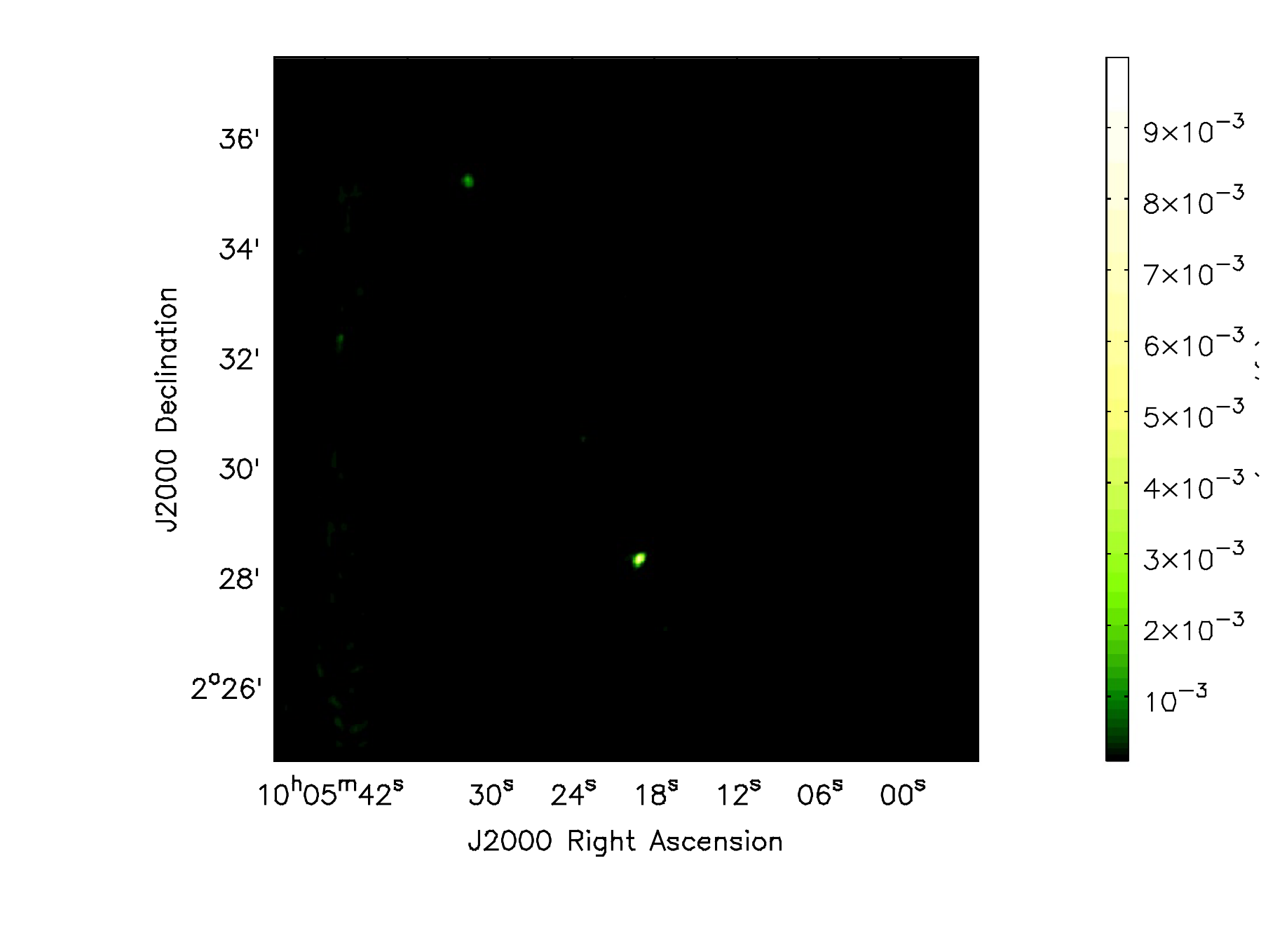}
    \includegraphics[width=0.3\textwidth]{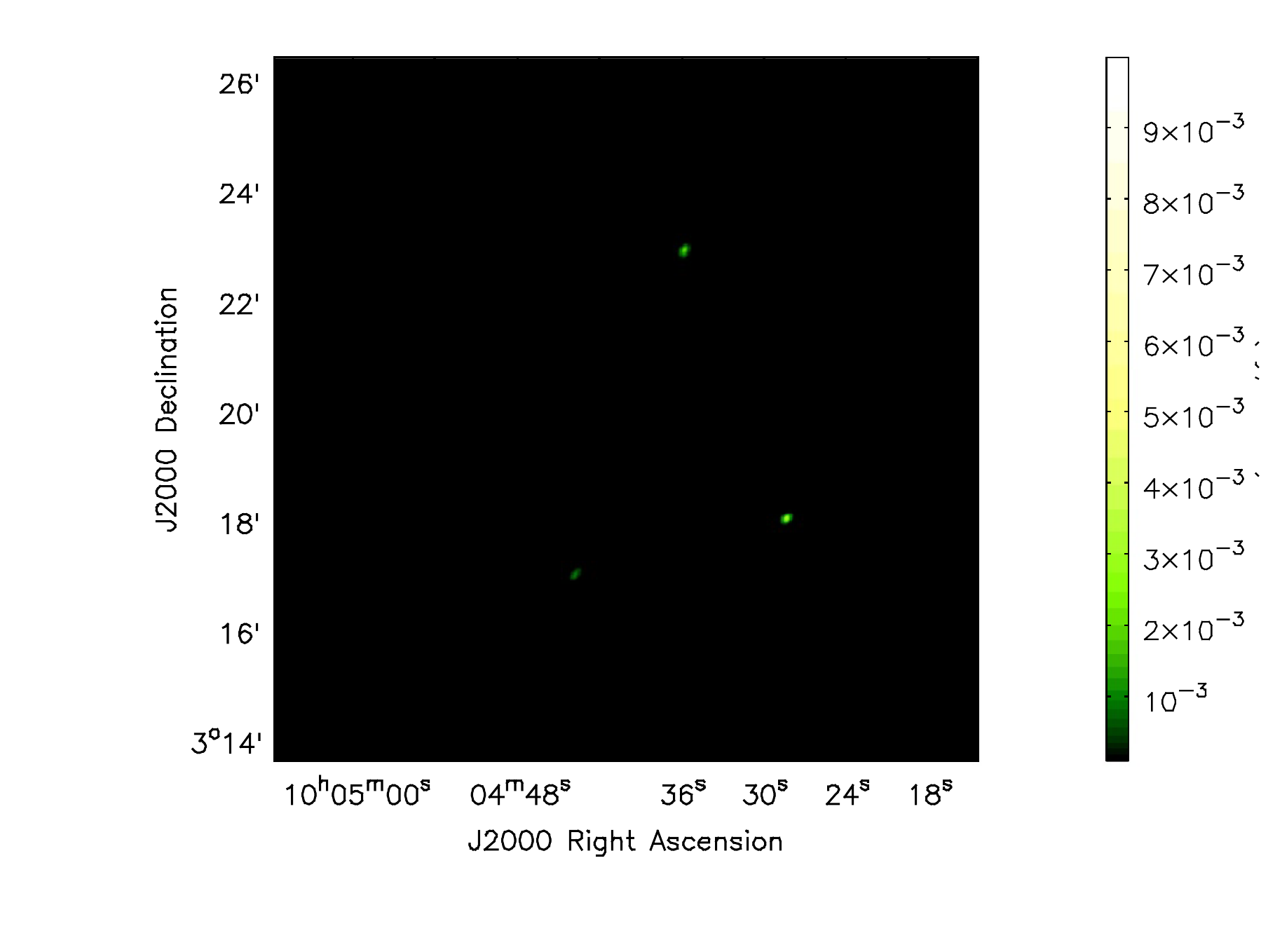}
    \includegraphics[width=0.3\textwidth]{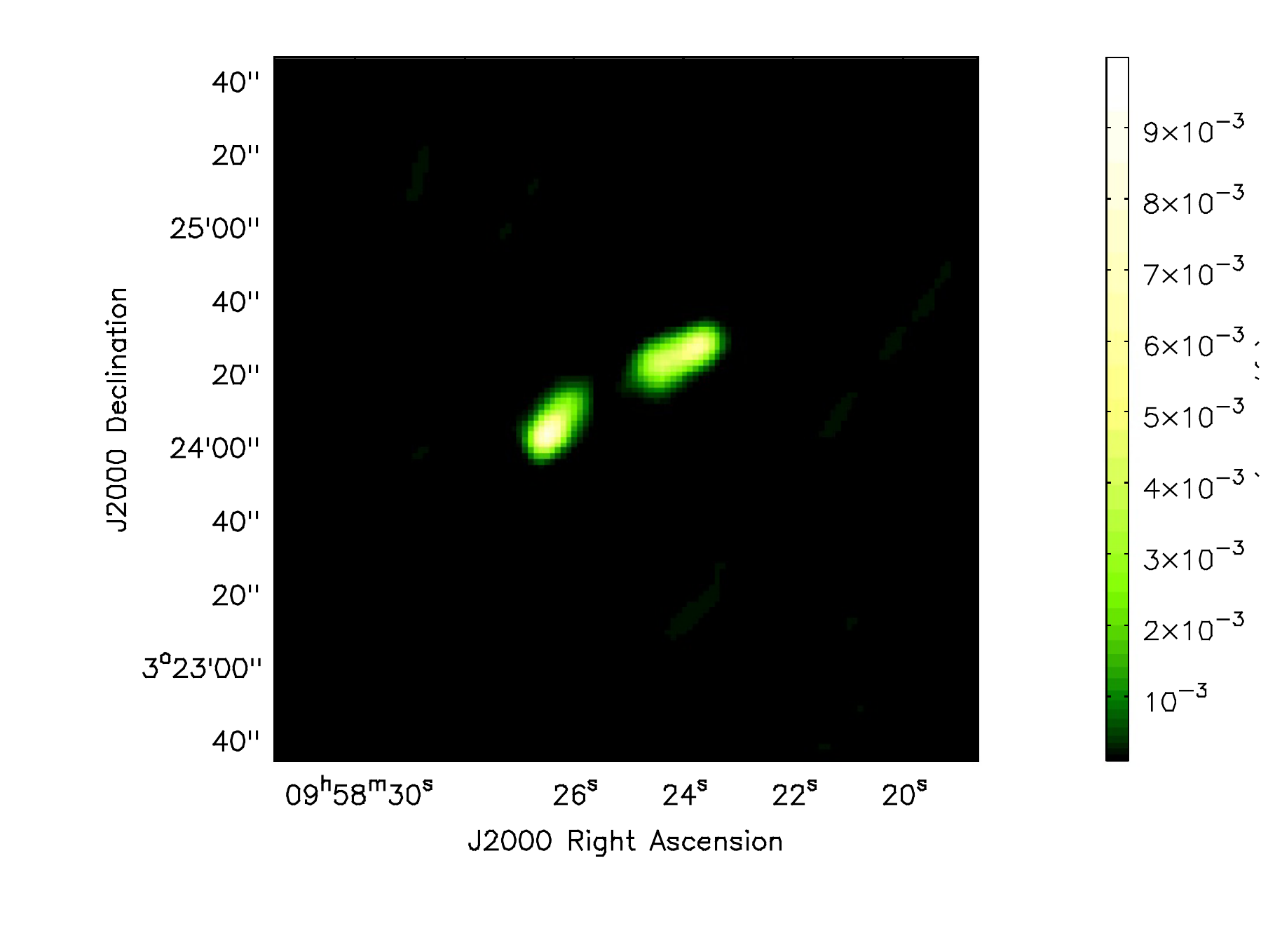}
    \includegraphics[width=0.3\textwidth]{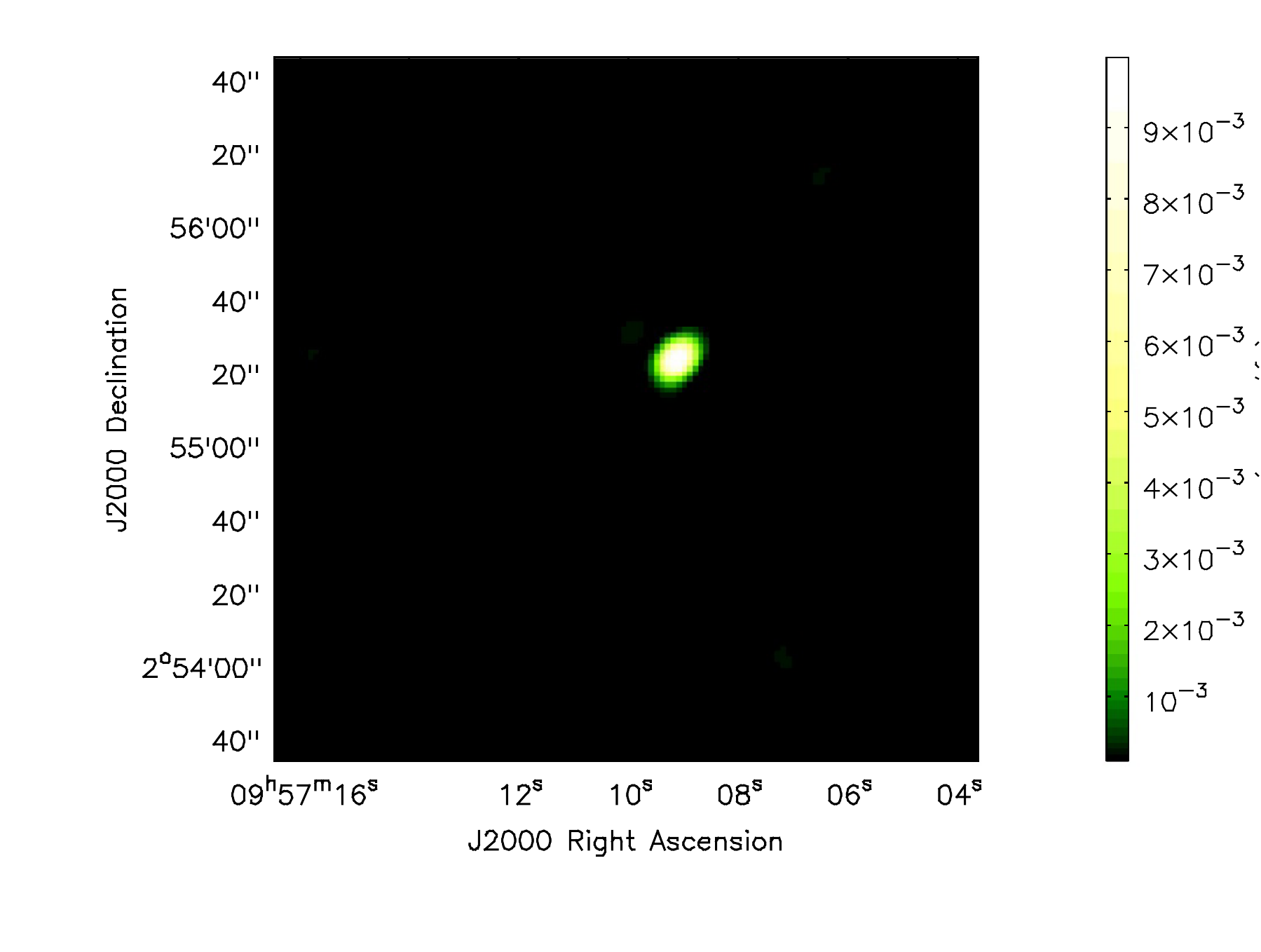}
    \includegraphics[width=0.3\textwidth]{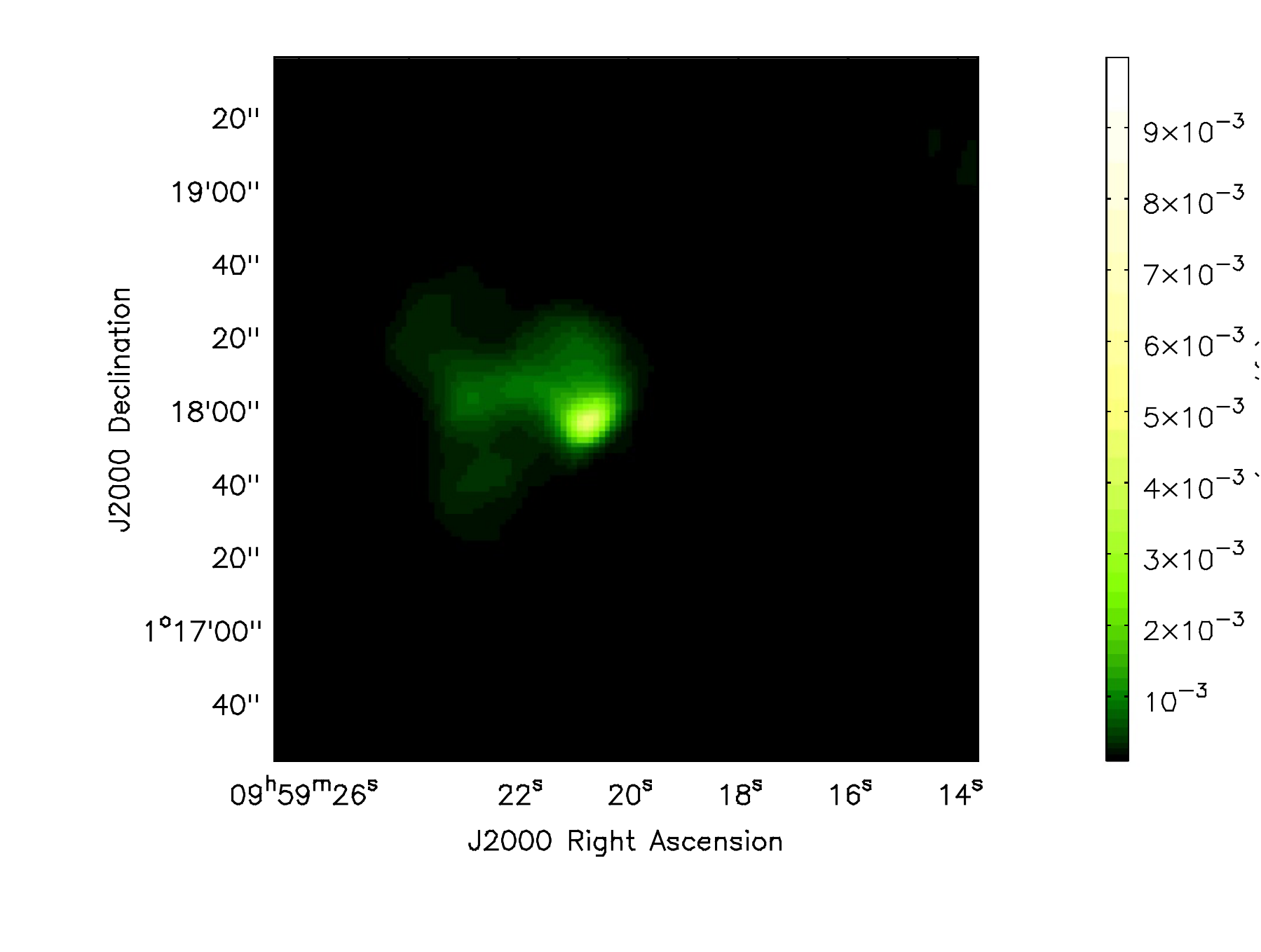}
    \includegraphics[width=0.3\textwidth]{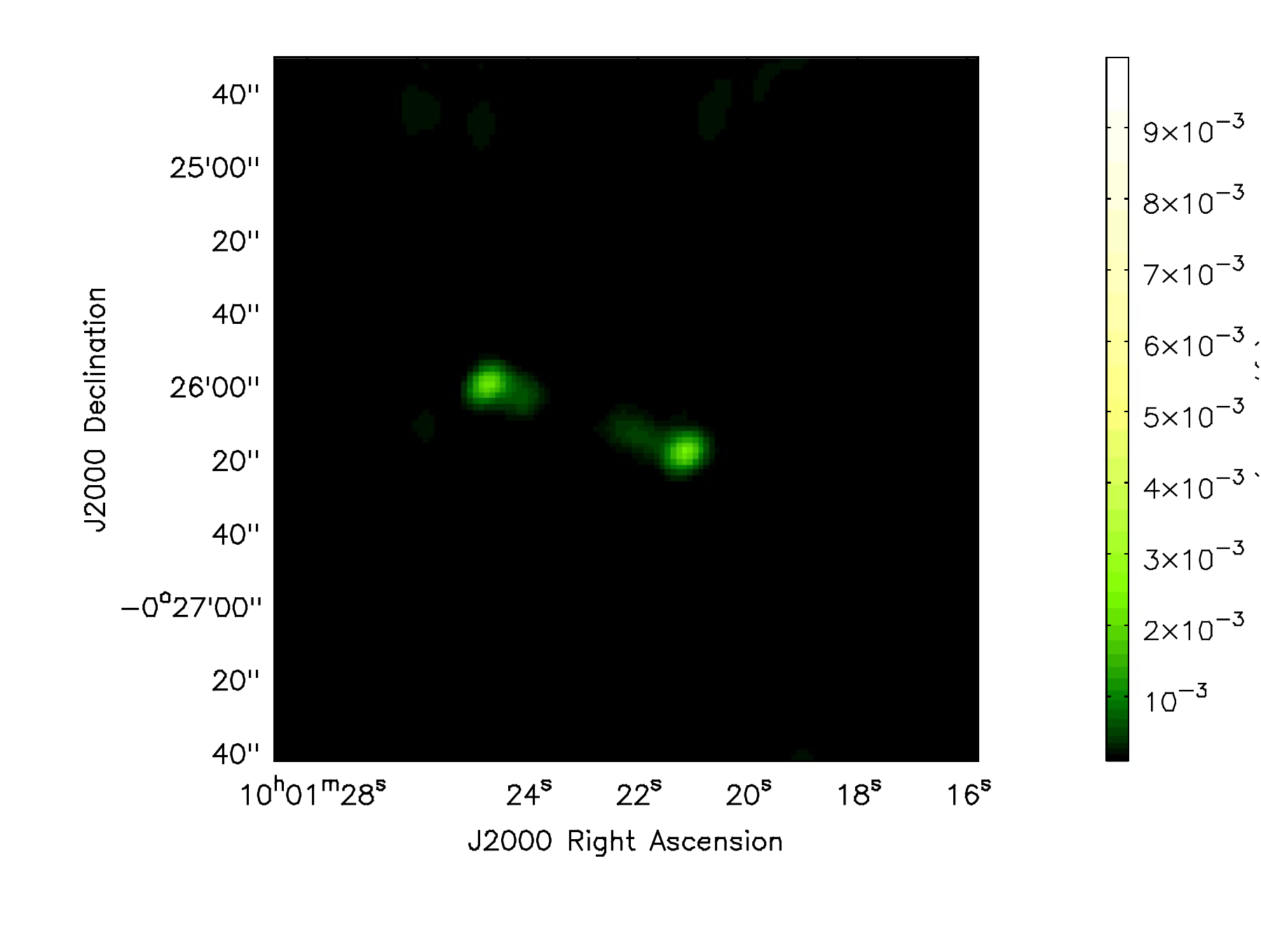}
    \includegraphics[width=0.3\textwidth]{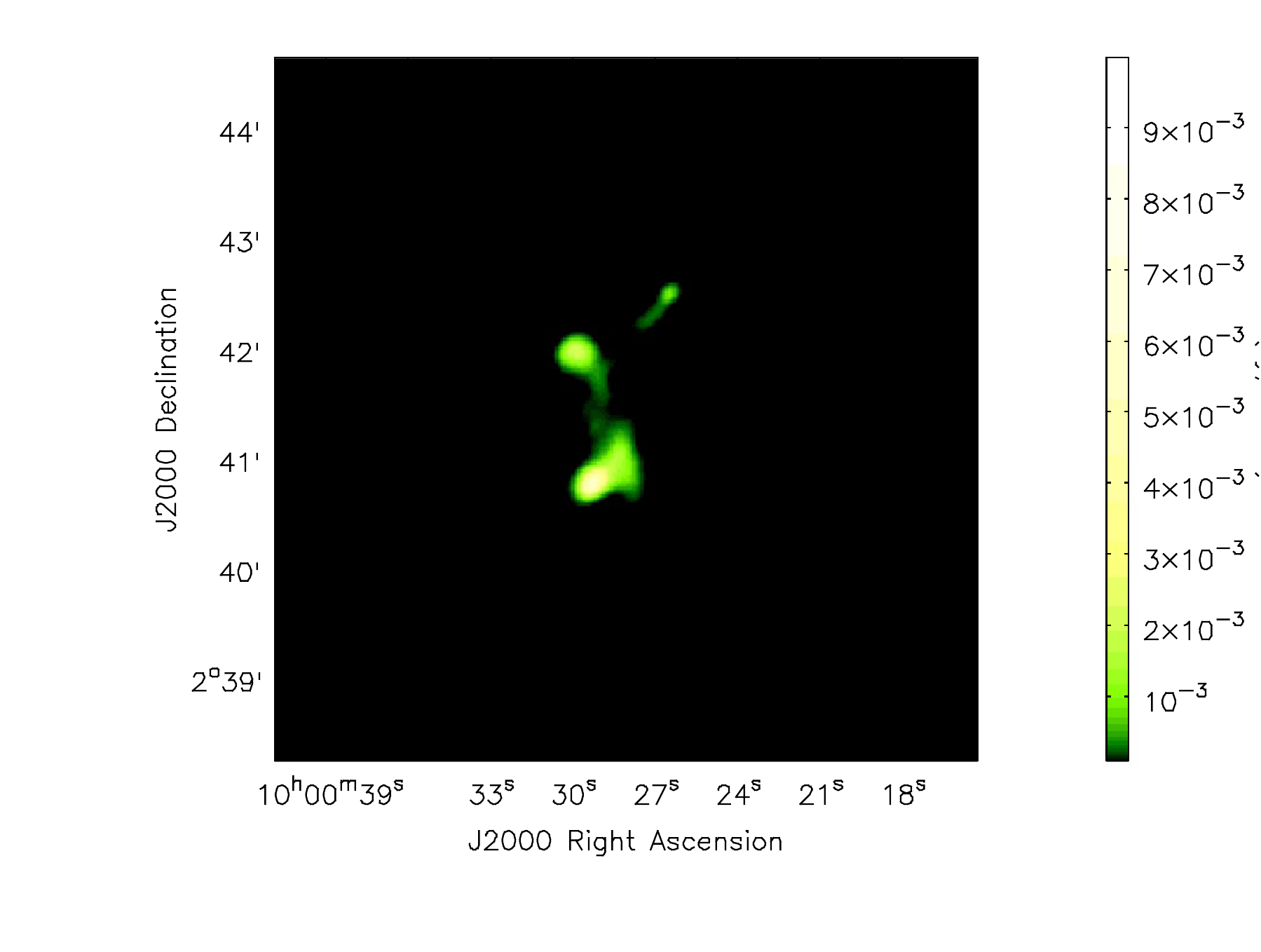}

    \caption{Continuum images of the eight strongest outliers to the CHILES field, from SPW 1. Channelised and HA binned models of these were subtracted from raw visibilities before imaging. The scale in Jy/beam is indicated by the color bar to the right of each panel.
    \label{fig:rogues}}
\end{figure}

Out-of-beam sources were the dominant cause of residual errors when the channel was not dominated by RFI, as shown in Fig. \ref{fig:in_beam}{\em c} and {\em d}. 
The contributions from a number of sources from outside the nominal field of view could  be detected as the frequency-dependent ripples that extended through out the preliminary cubes. These ripples have only a small effect on the RMS, but significantly inflate the maximum pixel values.
This highlights how important it is to have an accurate sky model extending far outside the field of view, which will be essential for SKA and the pathfinders.
It is illustrative to compare Fig. \ref{fig:in_beam}d with Fig. 4 in \citet[]{patil_16}; in their case these ripples were not seen, as their systematic contamination was not from residual sources. 
We settled on removing narrow-frequency continuum cube models of seven sources that by rights should have been undetectable (being well-beyond the nominal FoV) and one that was just beyond the half-power point. These are shown in Figure \ref{fig:rogues} and Table \ref{tab:rogues}. The one furthest from the FoV was at 10h01m24s -00d26m40s (4C -00.37) with a separation of about 2\degr{} from the telescope pointing. Nevertheless, as it was directly to the south of the phase center of the observations, the sidelobes from this source had some of the greatest impact on the image quality.

\begin{deluxetable*}{|l|cccr|}
\tablecaption{Key details of the eight strongest outliers removed from the CHILES field, along with the flux density cataloged in NED and the angular distance from the phase center. Models marked with a $\dagger$ are made up of multiple sources and the source name given is for the dominant component. \label{tab:rogues}}
\tablewidth{0pt}
\tablehead{
\colhead{Name} & \colhead{RA} & \colhead{Declination} & \colhead{Flux Density (Jy)} &
\colhead{Distance (\degr{})}}
\startdata
    WISEA J100314.86+012121.3 & 10h03m14.92s & +01d21m23.9s & 0.28 & 1.1 \\ 
    WISEA J100519.14+022813.6$^\dagger$ & 10h05m19.17s & +02d28m13.7s & 0.22 & 1.0 \\ 
    WISEA J100436.02+032248.8$^\dagger$ & 10h04m36.02s & +03d22m48.6s & 0.06 & 1.3 \\ 
    4C +03.17 & 09h58m26.32s & +03d24m01.5s & 0.29 & 1.3 \\  
    PMN J0957+0254 & 09h57m09.15s & +02d55m18.7s & 0.22 & 1.2 \\ 
    PKS J0959+0117 & 09h59m21.60s & +01d18m01.2s & 0.33 & 1.2 \\ 
    4C -00.37 & 10h01m23.06s & -00d26m09.5s & 0.85 & 1.9 \\  
    J150.12027+02.68098 & 10h00m28.87s & +02d40m51.5s& 0.05 & 0.4\\ 
\enddata
\end{deluxetable*}

\begin{figure}
    \centering
    \setlength{\unitlength}{\textwidth}
    \begin{picture}(0.5,0.35)
    \put(0,0){\includegraphics[width=0.5\textwidth]{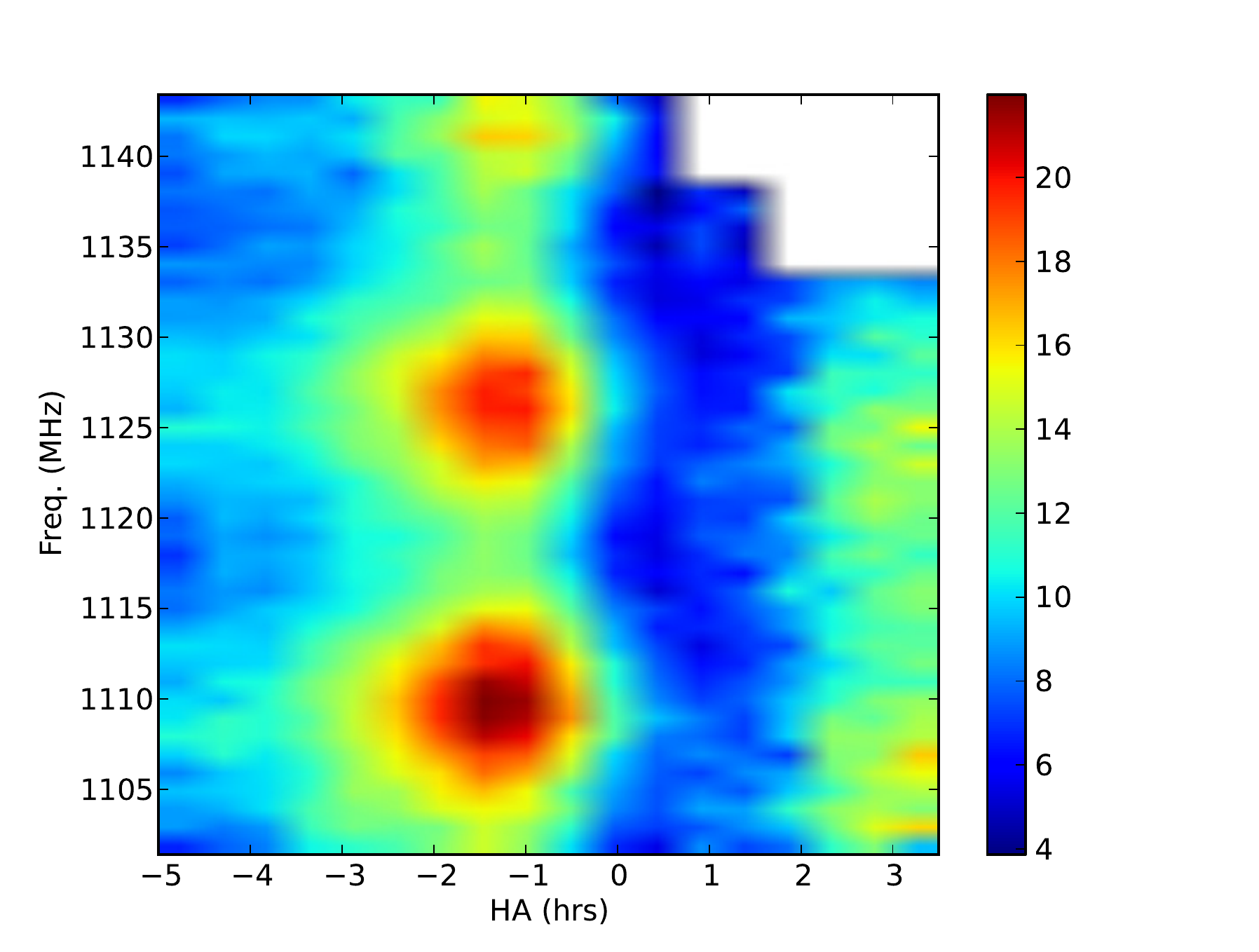}}
    \put(0.45,0.15){\rotatebox{90}{mJy/Pixel}}
    \end{picture}
    \caption{The maximum pixel value for the models of WISE\,J100314.86+012121.3 for SPW 5, covering 42\,MHz and 9 hours in Hour Angle. This source, being a point source, provides a clear example of the spectral and temporal behavior of the VLA instrumental response for sources so far outside the field of view. The top right corner does not contain any data, as these (short) observations happened to have the same frequency dither.}
    \label{fig:model_ha_freq}
\end{figure}

The in-beam field is corrected for frequency-dependent gain in the band-pass calibration, but that is not necessarily correct for sources far from the phase center, such as those that are dealt with in this step.
Therefore we generated our out-of-beam source models from the full first Epoch of observations (i.e. $\sim$200 hours), without a spectral index, at 1\,MHz steps. 
These sources are largely picked up in the sidelobes, which for the VLA are four lobes, arising from the feed support structure, which surround the main beam response \citep{vla_beam}. 
The projection of the support structure onto the sky rotates with the parallatic angle, so varies with Hour Angle (HA).
To account for the gain variation with HA for the outlier sources we binned the data into 1/2 hour steps and solved for an independent model in each of these steps. {The frequency and time averages were to provide sufficient SNR for modelling the sources, whilst being sufficiently fine to track the changing signals, as indicated in Figure \ref{fig:model_ha_freq}.
Various other combinations were trialed; of larger frequency spans and without accounting for time variation in the model. These did not reach acceptable levels of residuals. 
However, we note that another team in the CHILES collaboration is developing an alternative imaging pipeline, which subtracts daily models depending on these to account for the variation in LST coverage. This approach is delivering similar results.}
The models were checked for failed deconvolution, which was usually due to either RFI or limited data, or both. In these cases models were interpolated from nearby good channels. 
Figure \ref{fig:model_ha_freq} shows the peak flux for the first model, as a function of HA and frequency, for one SPW.

\begin{figure}
    \centering
    \plotone{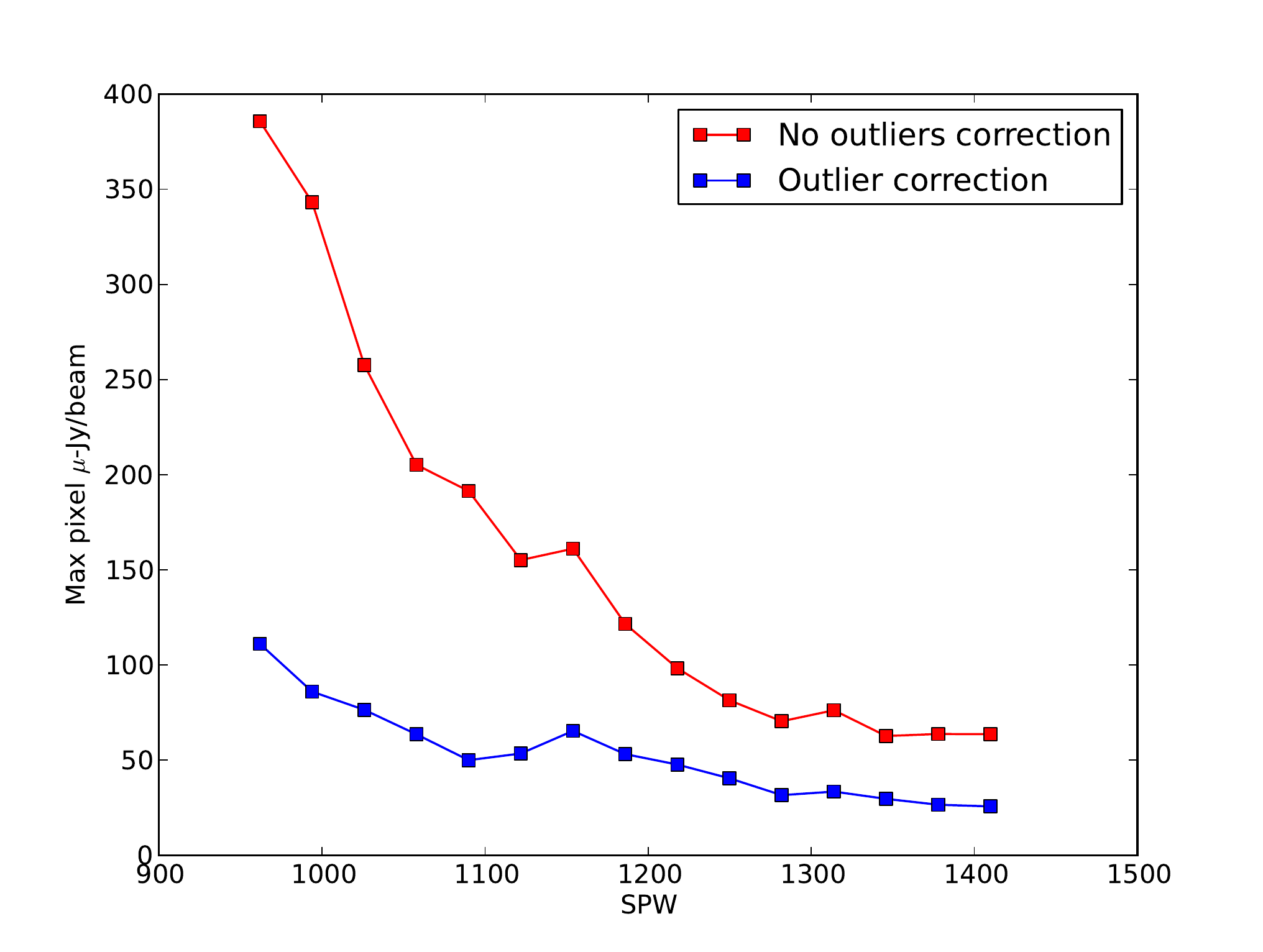}
    \caption{Peak residual flux in the images, with and without the subtraction of the out-lying sources, for all spectral windows. The peak residuals are much more sensitive to the sidelobes from the few isolated outlier sources than the RMS. 
    \label{fig:outeffect}}
\end{figure}

As these models were made from data phase rotated to their sky location, the model subtraction also had to be on data rotated to that line of sight, which was then rotated back to the phase center. This was done iteratively for each outlier source. 
After the outlying sources have been subtracted the images appear significantly improved, as shown in Fig. \ref{fig:in_beam} {\em c}. 
We plot the radial maximum absolute of the FFT of the images in Fig. \ref{fig:in_beam} {\em a-c} in Fig. \ref{fig:in_beam} {\em d}, to clearly show the suppression of the ripples after the outlier subtraction, which were not removed by the in-beam models. 
The RMS values are also lower, but the significant impact is on the ripples, as measured by the peak intensity pixel value. These are plotted for all SPW in Fig. \ref{fig:outeffect}.

\subsubsection{Gain Errors}
We believe that station-based gain errors (i.e. errors associated with instrumental effects from the individual antennas) would be the largest remaining residual contribution to the systematic errors. But these, as we show below, are below the thermal-noise limits in the spectral cubes. We are planning to investigate methods to correct for this when forming extremely deep continuum images. These corrections may be applied to the final data product that includes all the Epochs.
We have phase self-calibrated the data against the in-beam model, after subtracting the outlier models, with a phase solution per SPW per polarization per station every 30\,minutes. 
This produced phase solutions with a zero mean and a typical RMS of $\pm$5\degr{}, and was dominated by the measurement noise. 
As applying these solutions to the data would only increase the noise we have not used these in the cube. Future cubes will have them applied, if they improve the residuals.

We have amplitude self-calibrated the same data to solve for `un-normalised amplitude', with a solution per SPW per polarization per station for every observation.
This was to identify variations in the absolute gain scale over the few months of observations in Epoch 1. 
No significant variations could be identified, but there are suspicions that this will be an issue over longer timescales. 
{However, the CHILES VERDES search for transients \citep{chiles_verdes} showed only a few sources exhibited  significant variation. Even those which varied would either be only moderately variable or intrinsically weak.
Based on these results we would not expect the sky model to vary significantly over the observations.}

\subsection{Statistical measures from the cubes}
\label{sec:stats}

After subtracting the very best model that we could prepare, imaging with a limited number of clean iterations, and subtracting the residual continuum from the image cubes, we produce the final data product; a \HI{} line emission cube with continuous frequency coverage between 946 and 1420\,MHz in 0.25MHz steps. That is a redshift range of 0.5 to 0.0, with a velocity resolution of 50\,\kms{} at the highest frequency and 75\,\kms{} at the lowest.

\begin{figure}
    \centering
    \plotone{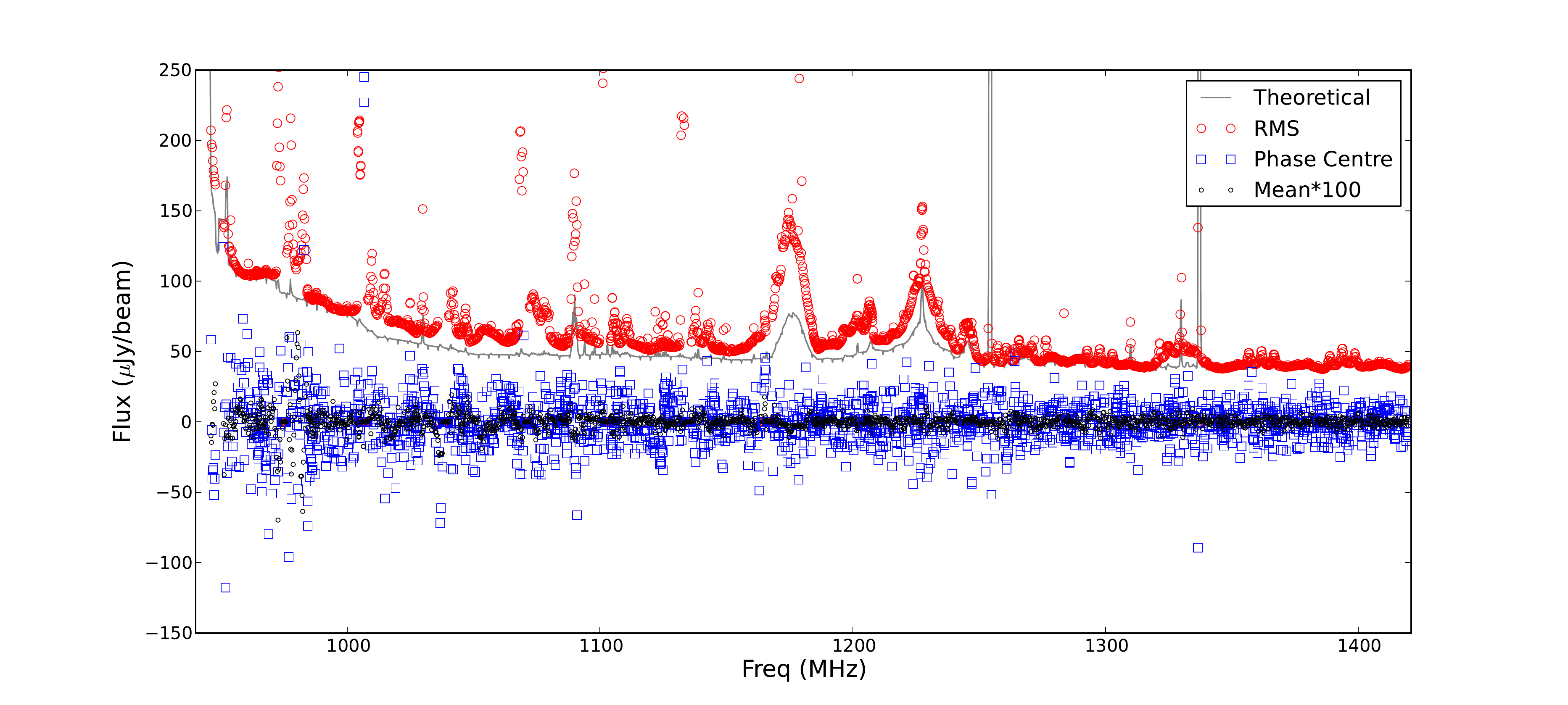}
    \caption{The per channel image cube RMS noise (red circles), the mean over the full Fov (scaled up by 100) (black dots) and the phase center profile (blue squares). This shows that the performance of the VLA remains excellent over the whole frequency band, despite the challenges from RFI and deep imaging. RFI residuals remain in those limited range of channels where the RMS is sharply elevated locally. For the vast majority of the channels we are close to matching the theoretical performance (gray line), based on the number of unflagged samples in each channel and the measured System Equivalent Flux Density.
    \label{fig:rms_dr2}}
\end{figure}

\begin{figure}
    \centering
    \plotone{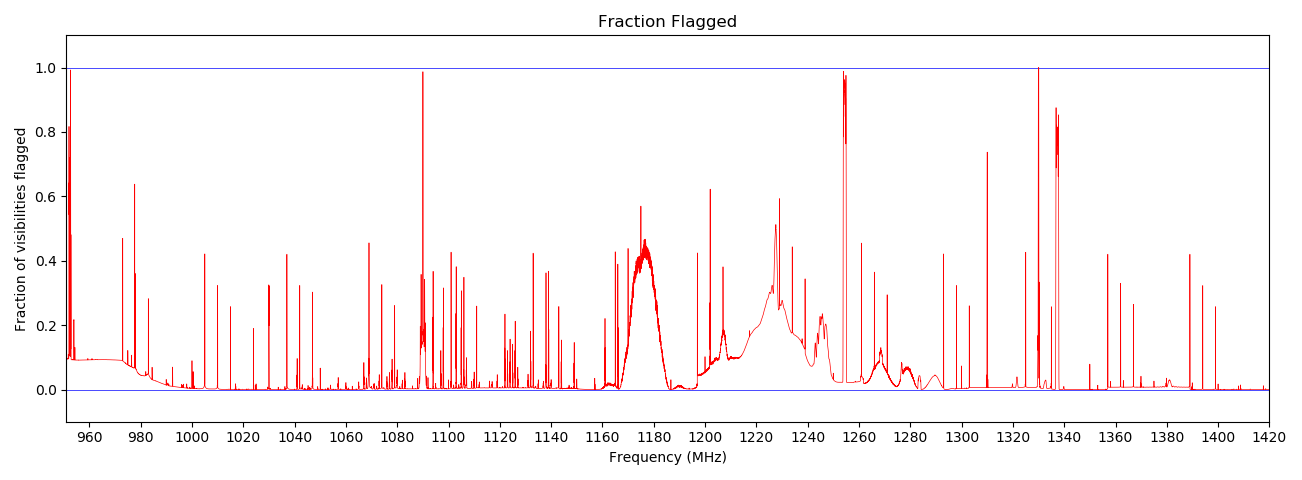}
    \caption{A plot of the fraction of flagged data as a function of frequency for all of our Epoch 1 observations of the CHILES field.  These flags exclude all online flagging (including bad antennas and data taken when not on source), and only include flags created by clipping, the {\sc rflag} algorithm, extending the flags, and any manual flags. The data loss at the band edges is noticeable in the triplets of lines at 32MHz spacing.}
    \label{fig:rfi_flags}
\end{figure}

We measured the RMS noise across the image for each frequency channel, as plotted in Fig. \ref{fig:rms_dr2}. In the absence of any residual non-Gaussian distribution, which was (finally) achieved by the subtraction of the out-of-beam sources, the RMS noise is the standard and best metric for the data quality.
The RMS noise is in close agreement with the expected VLA performance, based on the number of visibilities imaged in each channel. 
The frequencies where there are significant divergence are associated with known sources of RFI (as indicated in Fig. \ref{fig:rfi_flags}), underlying that we should be able to improve our flagging in future data products. However, increased flagging will reduce the amount of data and therefore increase the expected noise-levels. There will therefore be a limited return from these improved flagging strategies. 

The greatest challenge in the final CHILES image cubes will be to reach the thermal-noise limits. To test and confirm that we are on track to achieve this we have investigated the measured residual RMS noise in the cubes both as a function of the averaged bandwidth and time.

\subsubsection{Residual RMS noise as a function of Bandwdith}
\label{sec:resid_ch}

\begin{figure}
	\includegraphics[width=0.5\columnwidth]{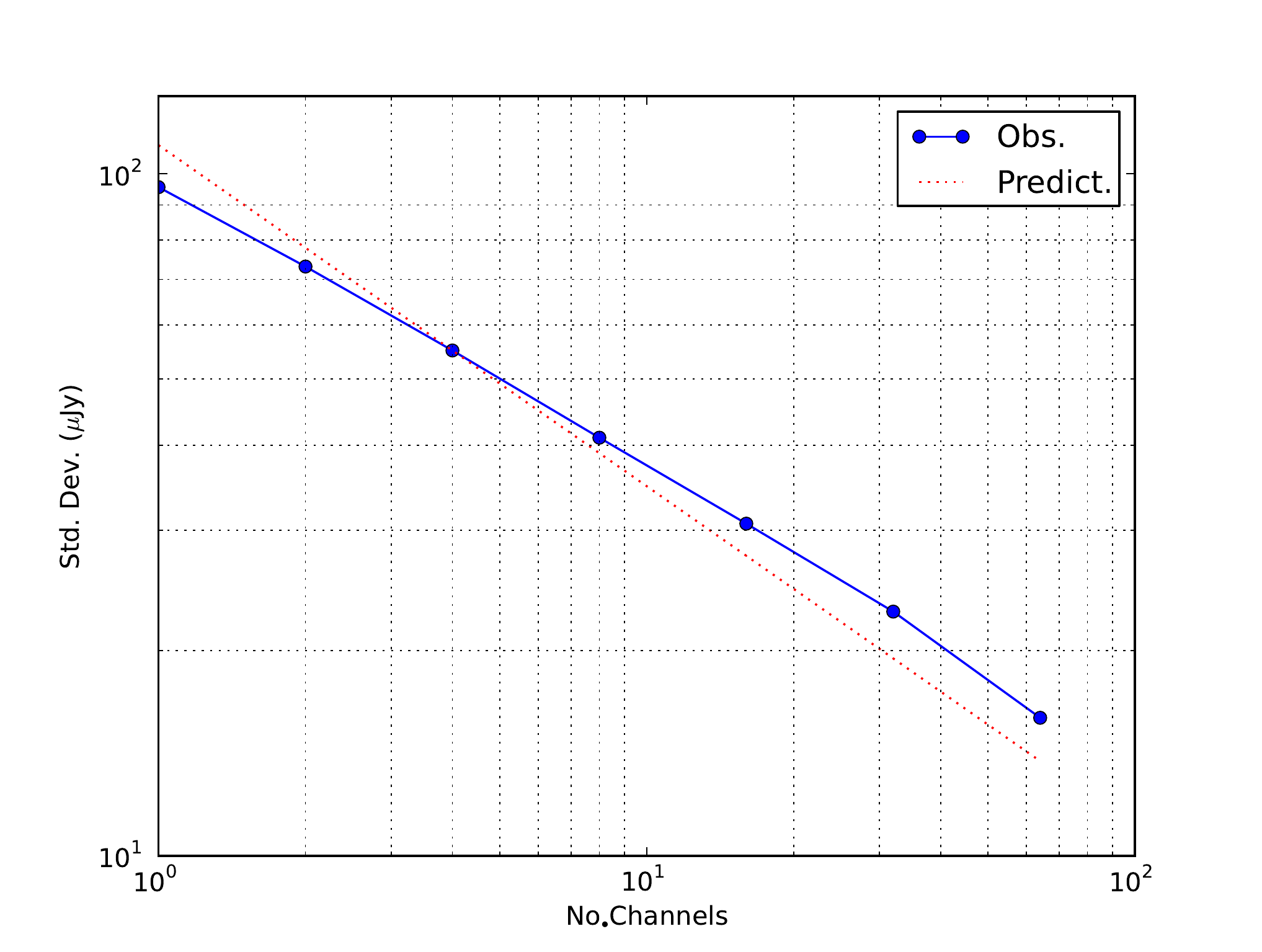}
	\includegraphics[width=0.5\columnwidth]{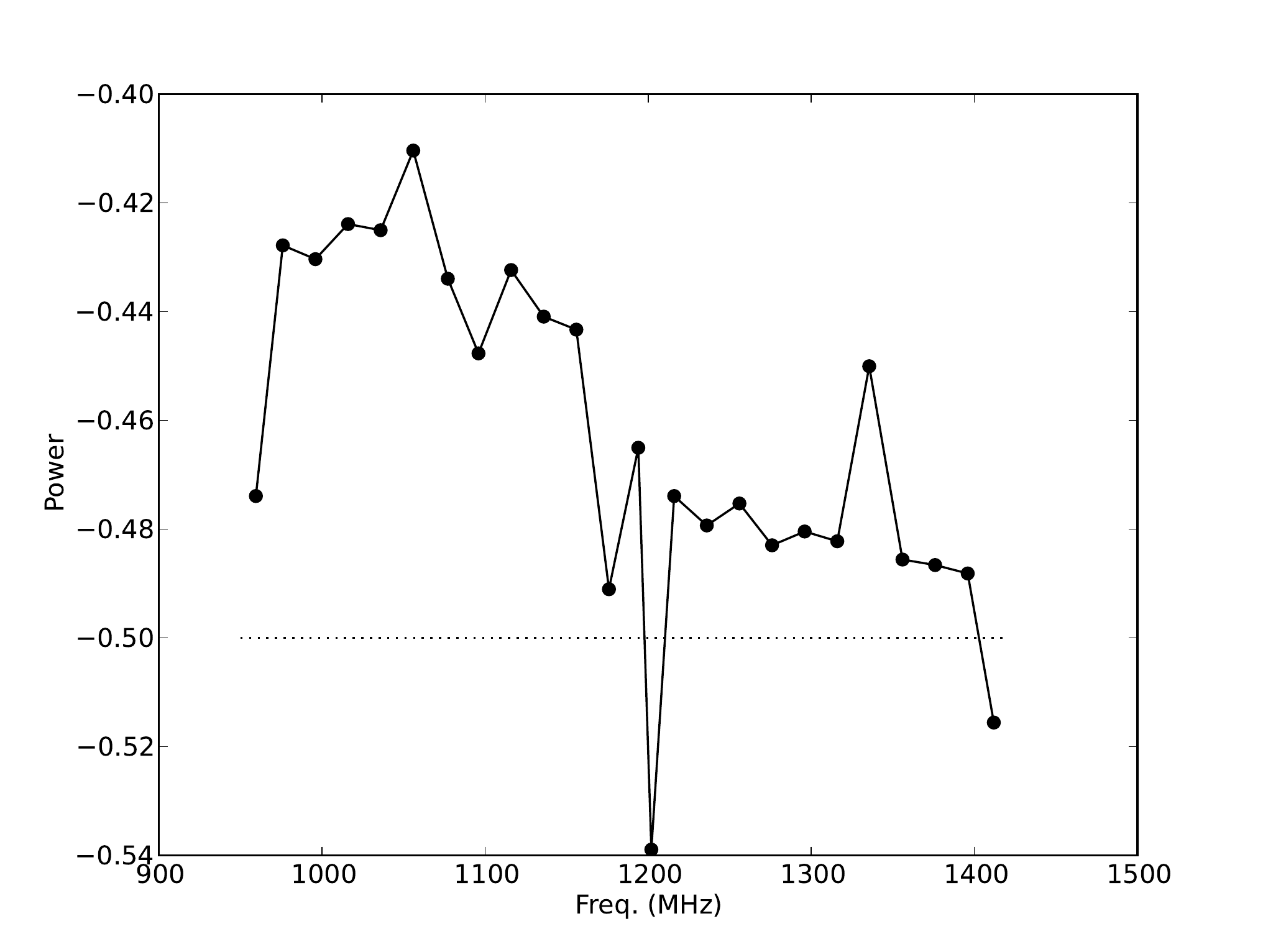}
    \caption{{\em left)} Mean residual RMS noise in a cube made with 62.5\,kHz channels between 1104 and 1128\,MHz, as a function of averaging over bandwidth. The red dotted line is the expected extrapolation for the RMS noise with the channel averaging in this cube (250\,kHz) with the power law index of -0.5, the expectation in the case of ideal uncorrelated noise. The best power law fit to the RMS noise is -0.45.
    {\em right)} The measured power law index for every 24\,MHz image cube, as a function of the averaging of channels. 
    \label{fig:rms_chan}}
\end{figure}

To measure the effect of the integration of multiple channels on the final image noise we formed images with channel averaging of 62.5 to 4,000\,kHz, from a 24\,MHz image cube with  62.5\,kHz channel-widths, made in the same fashion as our 250\,kHz channel-width image cube.
%
We plot the standard deviation (to avoid any issues with residual continuum sources) as a function of integration in Fig. \ref{fig:rms_chan}.
On the left it shows the residual noise from SPW 5 averaging down as a function of the number of combined channels, and the expected ideal behavior with a power law index of -0.5, interpolated from a channel width of 250\,kHz. On the right the power index measured in the same way is shown for all the SPW. In nearly all cases they fall slightly short of perfect behavior, with a typical index of -0.48. Nevertheless we consider this acceptable.

\subsubsection{Residual RMS noise as a function of integration time}\label{sec:resid_day}

We have confirmed that averaging multiple days together reduces the residual RMS as a function of the number of visibility records combined. This test was performed over the full spectral window of 32\,MHz, for spectral window 1 and 5 (centered on 994 and 1122 MHz respectively) on the uv-subtracted data. For these frequencies the expected confusion noise is less than 1 $\mu$Jy/beam \citep{condon_98}, so is below the thermal noise. 
The number of clean iterations was increased from 200 to 10,000 and no mask was used, to ensure that as much of the continuum flux was removed as was feasible.
\begin{figure}
	\includegraphics[width=0.5\columnwidth]{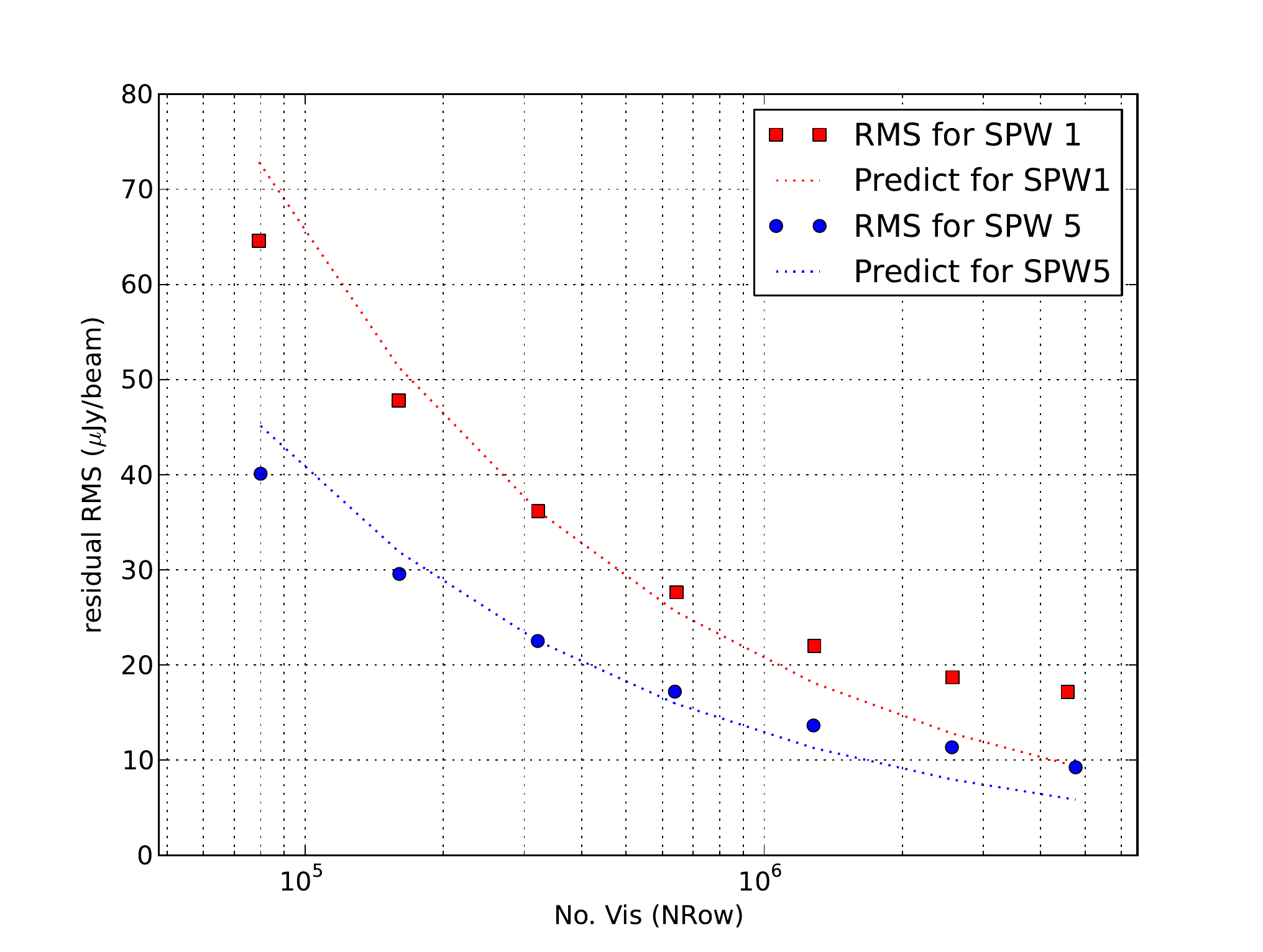}
	\includegraphics[width=0.5\columnwidth]{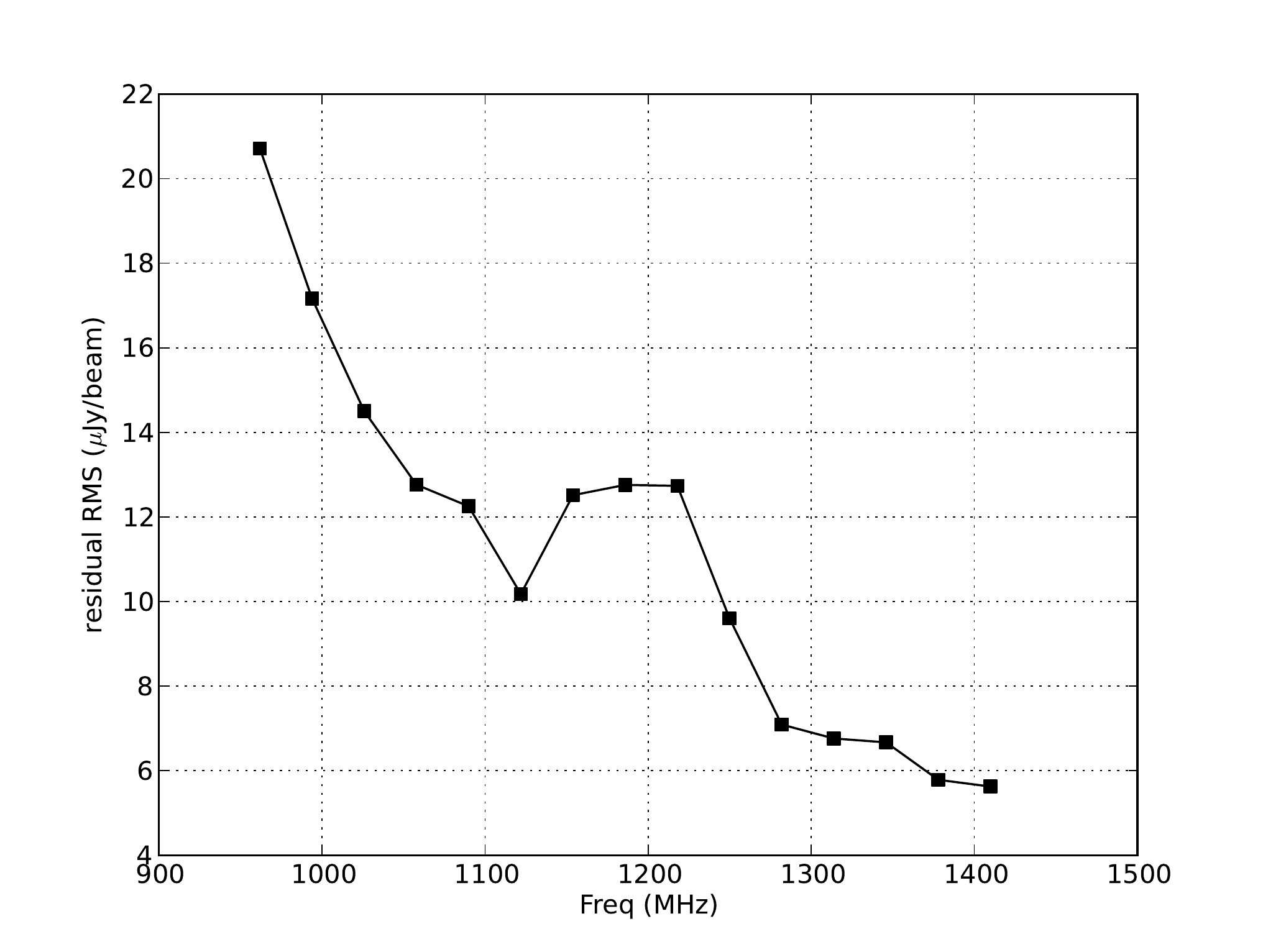}
    \caption{{\em Left}) Residual RMS noise as a function of the number of visibilities for the SPWs 1 and 5, at 994 and 1122\,MHz. This shows that we are 
    reaching a limit for the residual flux around $\sim$10$\mu$Jy, which is significantly above the confusion limit and about double the nominal thermal limits, but well below the expected spectral line noise. {\em Right}) Final residual RMS noise for all 32-MHz SPW, indicating the potential noise level for the complete 1,000\,hour observations should be $\sim$50$\mu$Jy per 250kHz channel. 
    \label{fig:rms_day}}
\end{figure}

From the uv-subtracted channel-averaged datasets for Epoch 1, we selected a number of scans (where a `scan' is the $\sim$10min observation of the target between calibrations) to form the final image. These were randomly selected from the dataset scan listing, and the residual image RMS is plotted against the number of visibility records involved in the imaging in Figure~\ref{fig:rms_day}.
On the left we show the residual noise from SPW 1 and 5 averaging down as a function of the number of visibilities combined, and the expected ideal behavior; that is the square root of the number of records scaled to the third data point. 
One can see that the residuals are approaching an asymptotic limit around 10$\mu$Jy for these two cases.
On the right we plot the limit of the residual RMS flux for all the SPWs.
These residuals are around double those predicted from the expected thermal limits, demonstrating that there are still systematic residuals in the data, but these are well below the level of the thermal limits expected for the 1,000 hour spectral cubes.


\section{Image Analysis}

We implemented a set of straight-forward Data Quality tests: 
i) a spectrum of the phase centre flux of the input visibilities, 
ii) a spectrum of a known in-beam continuum source in the image cube, 
iii) a spectrum of a known strong in-beam \HI{} galaxy in the image cube, 
iv) a frequency averaged 2D image, 
v) a declination averaged 2D image and 
vi) a statistical summary of the visibilities and the image cube.
These would: 
i) catch many of the data reduction, calibration, and/or imaging related errors. For example, when the uv-subtraction step failed the phase centre flux would be -1\,Jy; 
ii) confirm the successful subtraction of the in-beam spectral index model from the visibilities; 
iii) confirm the successful formation of a spectral image cube from the visibilities; 
iv) detect low signal to noise contamination, mainly out-of-beam continuum sources; 
v) detect RFI contamination of the images; 
vi) provide information for identification of the origin of any errors in the other tests. 
These data quality plots are prepared at the same time as the imaging, and are served from the same AWS website as for the image cubes themselves. 

\section{An Example Galaxy}

\begin{figure}
    \centering
    \includegraphics[width=0.6\columnwidth]{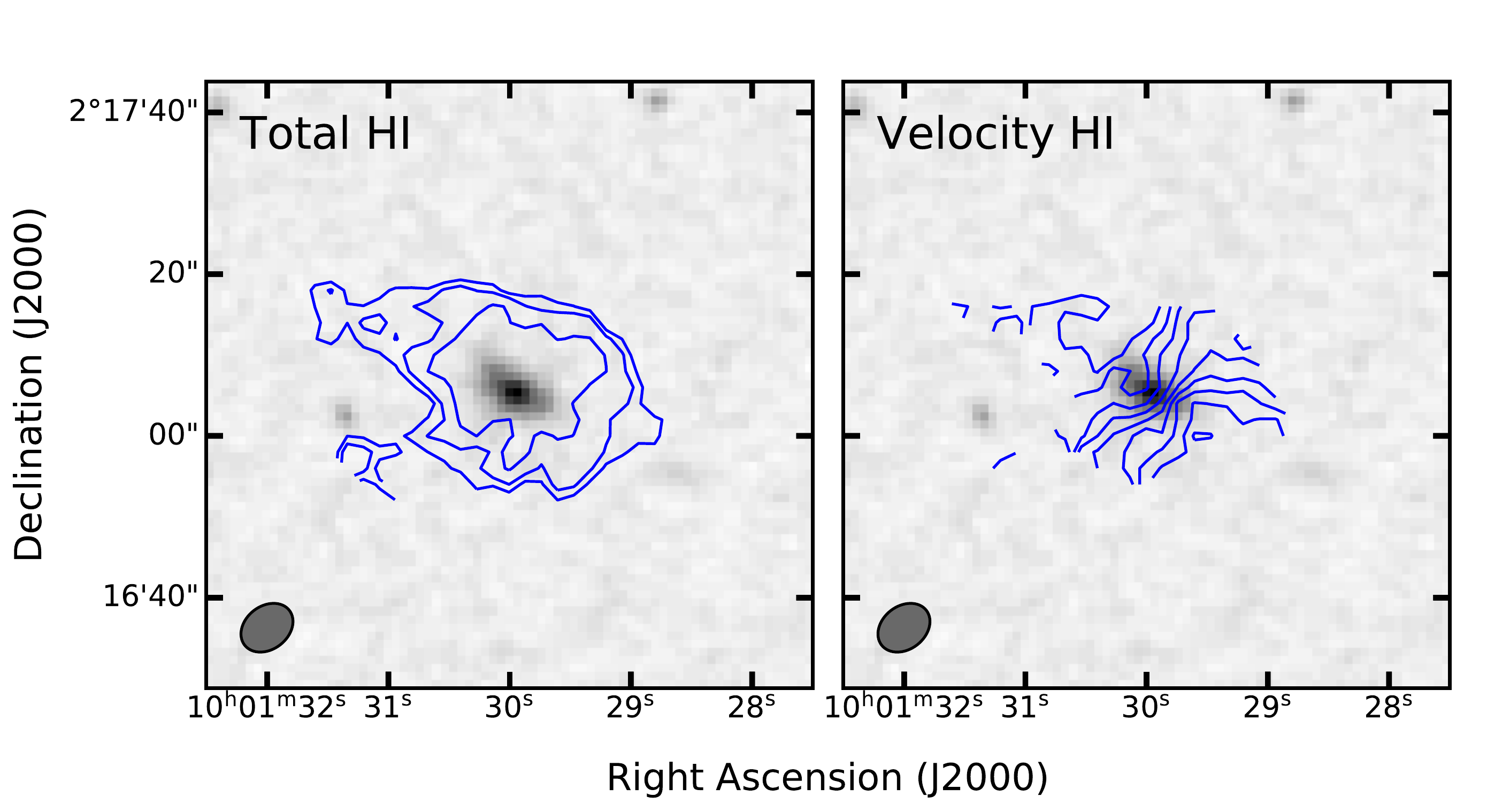}
    \includegraphics[width=0.3\columnwidth]{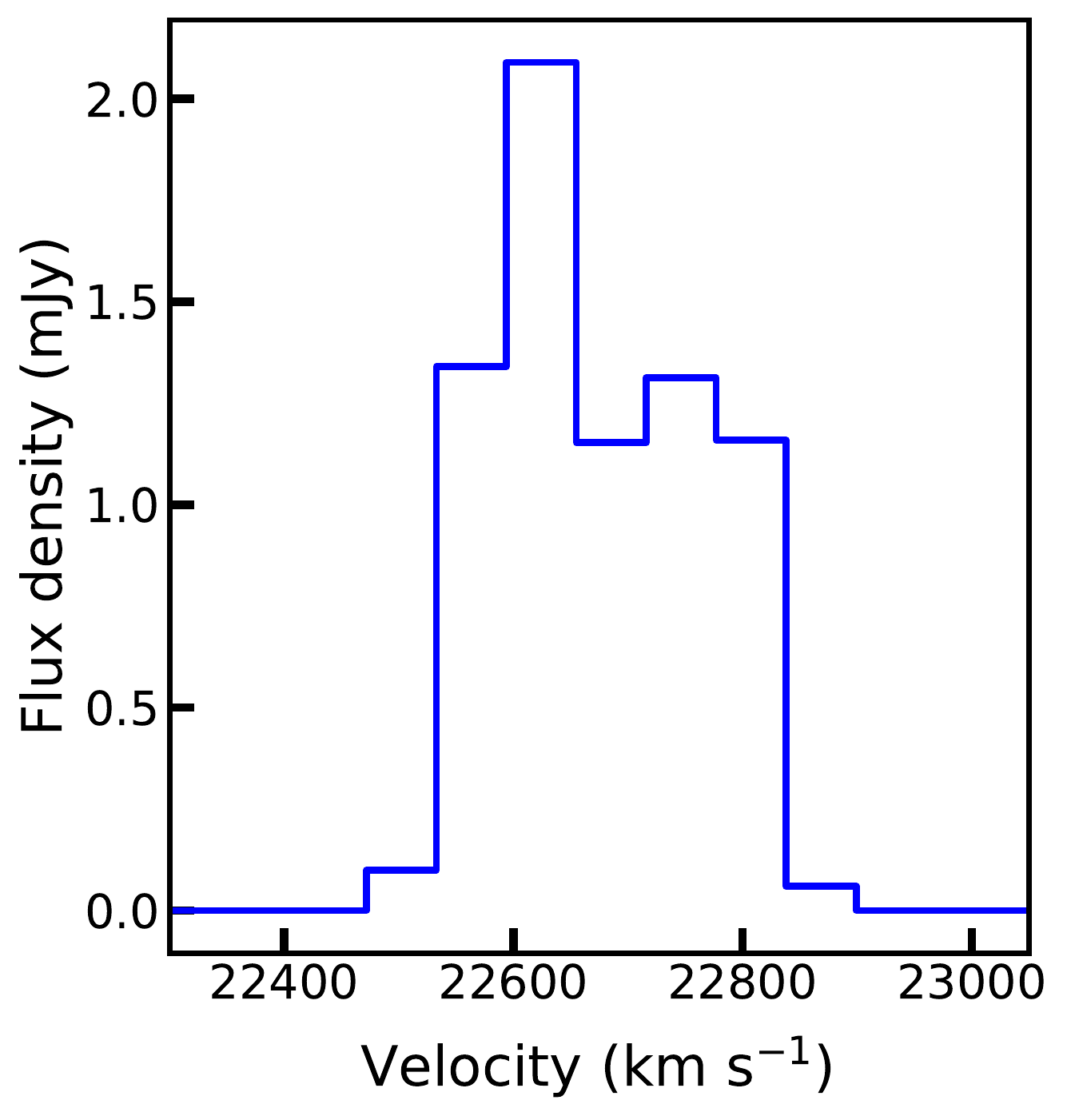}
    \caption{Properties for galaxy J100130.00+021705.0. Shown are from left to right the total intensity contours, the velocity field contours (both overplotted on the r-band DSS2 image) and the integrated spectrum. The total integrated flux contours are 3.5, 5 and 8$\sigma$. The contour values in cm$^{-2}$ are listed in the main text. The velocity field contours are given in $\pm$25\,km\,s$^{-1}$ increments from the selected central velocity of 22,676\,km\,s$^{-1}$. For both maps, the synthesized beam FWHM is shown in the bottom left, which is 7.1\arcsec{}$\times$5.3\arcsec{}}
    \label{fig:example_gal}
\end{figure}

As a demonstration of the data quality we have selected the known extended \HI{} galaxy at 1320MHz with an optical velocity of $\sim$22,600\,km\,s$^{-1}$, J100130.00+021705.0. The intensity and velocity maps alongside the integrated spectrum are shown in Figure~\ref{fig:example_gal}.  This source was also presented in \citet{chiles_local} where it is additionally labeled as 969633. This comparison allows us to confirm that these properties are consistent with our results, considering the difference in the two input data used in processing (i.e. an earlier version imaged at 62.5\,Hz per channel and the current version at 250\,kHz per channel). 

We used SoFiA \citep{serra_et_al_2015} to produce the integrated spectra, the total intensity and velocity field maps. We performed source finding on a $\pm$6MHz sub-band around the nominal position of the source, which had been separately deconvolved from the image cube, using the stored PSF, with the casa task {\sc deconvolve()}. 
For the source finding, the S\&C algorithm \citep{serra_et_al_2012} was used. The cube was spatially smoothed using Gaussian kernels of 4$\times$4 and 6$\times$6 pixels to detect emission at multiple resolutions. Furthermore, for each resolution a boxcar kernel with a channel width of three was used to smooth over the velocity axis. A relative flux threshold of 3.5$\sigma$ was used to detect significant pixels for each resolution. The final mask applied to the input cube was derived as the union of the masks found by SoFiA at each resolution.

The total intensity (moment 0) map and the velocity field (moment 1) map were overlaid as contours on r-band data from the 2\textsuperscript{nd} Digitized Sky Survey (DSS2) \citep{lasker_et_al_2000} obtained by using the skyview service of astroquery \citep{astroquery}. 
For the primary beam correction we used the standard parameters from {\sc PBCOR}, to give a factor of 0.96.
The integrated column density sensitivity is 
$\sigma\simeq$1.05 $\times$ 10$^{20}$ cm$^{-2}$ 
over $\sim$61\,km\,s$^{-1}$. The integrated flux contours shown are 
3.74,  5.27 and 8.43 
$\times$10$^{20}$ cm$^{-2}$. The central velocity for the contours was selected to be 22,676\,km\,s$^{-1}$ \citep{chiles_local}, and contours with $\pm$25\,km\,s$^{-1}$ increments are only shown where significant flux density ($\geq$3.5$\sigma$) is detected. The integrated spectrum is given by SoFiA as the sum of the flux density values of each pixel within the mask used. 
These results allow us to be confident that the image cubes are consistent with previous work and the image cubes are valid for use by the science teams.

\section{Conclusions}

We have demonstrated a data imaging pipeline for the {deepest-ever spectral line observation with the VLA, the} CHILES project, which sufficiently removes all detectable systematic errors to allow us to reach the thermal-noise limits from the first Epoch of observing. We are now confident that we can apply this to the remaining data and deliver a thermal-noise limited \HI{} cube with 1,000 hours of VLA observations. The analysis of the remaining four epochs is underway.
The challenges were, in the main, systematic-noise contributions from instrumental effects, underlining the importance of ensuring that instrumental responses are well designed, smooth and stable in future next-generation radio astronomy arrays.
{The unique challenge for CHILES was from sources far from the phase center, which could be detected because of the sensitivity, where there were uncorrected spectral and temporal instrumental effects.}
We were able to model the effects with piece-wise models generated from the data, segmented in frequency and hour-angle. 
Thus we were able to address these systematic errors, which were mainly in variations of the bandpass response across the field of view, particularly in regions outside those normally considered for instrument design.

When the visibilities are available these instrumental effects can be isolated and corrected for in the data. 
This would be difficult-to-impossible to do in the image domain. 
This therefore underlines the importance of retaining visibilities in some fashion for the proposed deep spectral line \HI{} and/or EOR surveys planned for the next-generation instruments \citep[e.g.][]{ngvla,ska_aas}.
This will have a major impact on the planning for deep spectral line surveys. Fortunately solutions for this have been proposed, either by storing the baseline-averaged visibility data \citep{wijnholds_18} or by storing them, with correct kernels applied, on the uv-grid \citep{evlamemo_198,kristof_deep}.

\section*{Acknowledgments}

The CHILES survey was partially supported by a collaborative research grant from the National Science Foundation under grant numbers AST - 1412843, 1412578, 1413102, 1413099 and 1412503.  

This research was supported by the Australian Research Council Centre of Excellence for All Sky Astrophysics in three Dimensions (ASTRO-3D), through project number CE170100013 and by grants from Amazon Web Services, the AstroCompute project.
%

The National Radio Astronomy Observatory is a facility of the National Science Foundation operated under cooperative agreement with Associated Universities Inc.. 
Source flux densities in Table \ref{tab:rogues} were taken from the NASA/IPAC Extragalactic Database (NED), which is funded by the National
Aeronautics and Space Administration and operated by the California Institute of
Technology.

J.M.vd.H. acknowledges support from the European Research Council under the European Union’s Seventh Framework Programme (FP/2007-2013)/ERC grant agreement no. 291531 (HIStoryNU).
D.J.P., N.L., and E.S. acknowledge partial support from NSF grants No. AST 1412578 and AST-1149491 and from the WVU Eberly College Dean's office.
K. R. acknowledges support from the Bundesministerium für Bildung und Forschung (BMBF) award 05A20WM4.

\section*{Data Availability Statement}
The raw data underlying this article are available in the NRAO archives. The processed data underlying this article are subject to an embargo of 12 months from the publication date of this article. Once the embargo expires the processed image cubes will be available via the project PI.


\end{document}